\documentclass[fontsize =12pt ,
paper =a4, 
draft =false , 
titlepage=false , 
twoside =false , 
open =right , 
parskip=half , 
DIV =12]{scrartcl}
\usepackage[utf8]{inputenc}
\usepackage[T1]{fontenc}
\usepackage[english]{babel}
\usepackage{amsmath}
\usepackage{dsfont}
\usepackage{amsfonts}
\usepackage{bm}
\usepackage{csquotes}
\usepackage{tikz}
\usepackage{booktabs}
\usepackage{array}
\usepackage{enumitem}
\usepackage{adjustbox}
\usepackage{multirow}
\usepackage{tabularx}
\usepackage{authblk}
\usepackage{hyperref}
\usepackage[defaultlines = 3, all]{nowidow}
\usepackage[dvipsnames]{xcolor}
\usepackage{subcaption}
\usepackage{titlesec}

\usepackage{setspace}
\usepackage{anysize}
\marginsize{2cm}{2.5cm}{2.5cm}{2cm}

\usepackage{hyperref}
\usepackage{url}
\usepackage{xurl}
\usepackage{graphicx}
\usepackage{amssymb}

\usepackage[backend=biber, 
			style=apa, 
			citestyle = authoryear-comp, 
			maxcitenames=2, 
			maxbibnames=20, 
			dashed = false, 
			giveninits=true, 
			url = true, 
			natbib=true, 
			doi = false, 
			eprint =false, 
			bibstyle =authoryear, 
			isbn=false, clearlang=true, 
            uniquename=false,
			uniquelist=false]{biblatex}

\renewbibmacro*{volume+number+eid}{
	\printfield{volume}
	\setunit*{\addnbthinspace}
	\printfield{number}
    \setunit{\addcomma\space}
	\printfield{eid}}
\DeclareFieldFormat[article]{number}{\mkbibparens{#1}}
\DeclareFieldFormat[article]{eid}{#1}

\AtEveryBibitem{\clearfield{month}}
\AtEveryCitekey{\clearfield{month}}
\AtEveryBibitem{\clearfield{day}}
\AtEveryCitekey{\clearfield{day}}
\AtEveryBibitem{\clearfield{note}}
\AtEveryCitekey{\clearfield{note}}

\providecommand{\keywords}[1]
{
  \small	
  \textbf{\textit{Keywords---}} #1
}

\NewBibliographyString{toappear}
\DefineBibliographyStrings{english}{
  toappear = {to appear in:},
  version = {R package version},
}

\renewbibmacro*{in:}{
  \iffieldundef{pubstate}
    {}
    {\printfield{pubstate}
     \setunit{\addspace}
     \clearfield{pubstate}}
    }

\usepackage{breakurl}

\newcolumntype{C}[1]{>{\centering\let\newline\\\arraybackslash\hspace{0pt}}m{#1}}

\usepackage{xr}
\usepackage{sectsty}
\usepackage{lmodern} 
\usepackage{relsize}
\sectionfont{\rmfamily\bfseries\normalsize} 
\subsectionfont{\rmfamily\bfseries\normalsize}

\begin{document}
    \title{ \vspace{-1em}\Large{\textrm{Parameter Estimation and Seasonal Modification of the Fractional Poisson Process with Application to Vorticity Extremes over the North Atlantic}}}
    \author[1, 2]{Merle Mendel}
    \author[1]{Roland Fried}
    \affil[1]{Department of Statistics, TU Dortmund University, Germany}
    \affil[2]{Corresponding author: merle.mendel@tu-dortmund.de}
	\maketitle

\begin{center}
    \rmfamily\large\bfseries Abstract
\end{center}
{\fontsize{11pt}{14pt}\selectfont The fractional Poisson process (FPP) generalizes the standard Poisson process by replacing exponentially distributed return times with Mittag-Leffler distributed ones with an extra tail parameter, allowing for greater flexibility. The FPP has been applied in various fields, such as modeling occurrences of extratropical cyclones in meteorology and solar flares in physics. We propose a new estimation method for the parameters of the FPP, based on minimizing the distance between the empirical and the theoretical distribution at selected quantiles. We conduct an extensive simulation study to evaluate the advantages and limitations of the new estimation method and to compare it with several competing estimators, some of which have not yet been examined in the Mittag-Leffler setting. To enhance the applicability of the FPP in real-world scenarios, particularly in meteorology, we propose a method for incorporating seasonality into the FPP through distance-based weighting. We then analyze the return times of relative vorticity extremes in the North Atlantic-European region using our seasonal modeling approach.}


	\section{Introduction}

The fractional Poisson process (FPP), introduced by \citet{fpp_original}, offers a natural generalization of the widely used standard Poisson process. It enables the description of temporal clustering of events, often found in real-world applications. \citet{hees2020} applied the FPP to solar flare data, \citet{blender} suggested its use for modeling extratropical winter cyclone occurrences, and \citet{Dissanayake} examined it for significant wave heights measured on the Sefton Coast (UK). In the FPP, the return times follow a Mittag-Leffler distribution instead of an exponential distribution, allowing for heavy-tailed behavior.  

The first estimation method proposed for the Mittag-Leffler parameters was the fractional moment estimator of \citet{kozubowski}, which requires selecting appropriate constants that depend on the true parameter values prior to the estimation. \citet{logmoment} addressed this limitation by introducing an estimator based on the first two moments of the logarithmized random variable. This method of logarithmic moments is one of two estimation approaches commonly used in the Mittag-Leffler setting nowadays, the other being maximum likelihood estimation \citep{hees2020}.

Using the Mittag-Leffler distribution to model waiting times between extreme events can be justified by asymptotical arguments. Under certain assumptions, the Mittag-Leffler distribution serves as an asymptotic approximation to the unknown true waiting time generating distribution \citep{hees2020}. Therefore, we are interested in finding an estimation approach with some robustness against stringent distributional assumptions, particularly concerning the tail behavior. \citet{christina} minimize the Cramér-von-Mises distance for parameter estimation in fractional compound Poisson processes. As this setting is similar to the FPP, we adopt this approach for Mittag-Leffler parameter estimation. 
Furthermore, we consider the quantile least squares estimator, suggested in \citet{QLS} for robust parameter estimation in location-scale families. Minimizing distances between the quantiles of the empirical and the model distribution aims at good approximations of distributional characteristics which are relevant in many applications. 
However, as we will see later on, the quantile least squares method has some problems in the Mittag-Leffler setup. Therefore, we propose a new quantile-based estimation method for the Mittag-Leffler distribution. It combines ideas underlying the Cramér-von-Mises distance and the quantile least squares approach and minimizes the sum of the squared distances between the empirical and the theoretical distribution function at selected quantiles. 
We compare the different estimation approaches for the Mittag-Leffler parameters in an extensive simulation study, covering various parameter settings and analyzing empirical relative efficiencies measured by mean squared errors, computation time, and robustness of the methods. 

In view of the strong seasonal patterns found, e.g., in the waiting times between extreme weather phenomena, we propose an approach to incorporate seasonality into the FPP by allowing the parameters of the return time distribution to vary periodically. We suggest estimating these time-dependent parameters using a distance-based weighting approach, with weights determined by an Epanechnikov kernel \citep{epanechnikov}, and a weighted version of the log-moment, maximum likelihood or quantile-based estimator.   
This approach enables flexible modeling of various forms of seasonality while imposing only minimal assumptions. We apply this method to extreme relative vorticities, which are commonly used in cyclone detection. \citet{blender} and \citet{christina} proposed the homogeneous FPP with constant parameters to model winter vorticity extremes. Our seasonal approach allows modeling the relative vorticity extremes throughout the entire year. 
The results indicate temporal clustering of the vorticity extremes in winter, as well as during most of spring and fall. In contrast, during summer the clustering is less pronounced, and a large fraction of the examined region exhibits a behavior as expected in an ordinary Poisson process.

The remainder of this paper is organized as follows. Section \ref{sec:methods1} reviews the existing methodology for fitting the Mittag-Leffler distribution and the homogeneous fractional Poisson process. We outline further estimators that may be of interest in the Mittag-Leffler context and introduce a new quantile-based estimation method for the Mittag-Leffler parameters. Furthermore, we suggest an approach for including seasonality in the FPP.
Section \ref{sec:simulation} provides a comprehensive simulation study that evaluates the previously discussed estimation procedures for the Mittag-Leffler distribution in different settings.
In Section \ref{sec:example}, we apply the fractional Poisson process with seasonal variation to data on relative vorticity extremes in the North Atlantic-European region and analyze the results, with a focus on spatial and seasonal differences.
Finally, we summarize and discuss our findings in \mbox{Section \ref{sec:conclusion}.}

\section{Methodology} \label{sec:methods1}

In this section, we review the existing estimation methods and suggest a new quantile-based estimation approach for the parameters of the Mittag-Leffler distribution. Furthermore, we extend the fractional Poisson process by incorporating time-varying parameters in the Mittag-Leffler distribution. We propose estimating these time-dependent parameters with a distance-based weighting approach, inspired by the weighted moving average.

\subsection{Quantile-Based Estimation in the Mittag-Leffler Distribution} \label{sec:methods_1}

The Mittag-Leffler distribution was introduced by \citet{mittagleffler_original}. The distribution gets its name from the Mittag-Leffler function $E_{\beta}(x)$, which is defined as
\[E_{\beta}(x) = \sum\limits_{k = 0}^\infty \frac{x^k}{\Gamma(1 + \beta k)}\]
with input $x\in \mathbb{C}$ and parameter $\beta \in \mathbb{C}$ with $\text{Re}(\beta) > 0$. A real-valued random variable $X$ is Mittag-Leffler distributed with tail parameter $\beta \in (0,1]$ and scale parameter $\sigma > 0$, if its cumulative distribution function (CDF) is given by
\[F_{\beta, \sigma}(x) = \left\{
\begin{array}{ll}
1 - E_{\beta}\left(-\left(x/\sigma \right)^{\beta}\right), &\quad x > 0,\\
0, &\quad x \leq 0.
\end{array}
\right.\]
For $\beta < 1$, the moments of a Mittag-Leffler distributed variable $X$ do not exist \citep{logmoment}.
In the special case of $\beta = 1$, the Mittag-Leffler distribution reduces to the exponential distribution with rate parameter $\lambda = \frac{1}{\sigma}$.
Hence, the Mittag-Leffler distribution can be seen as a generalization of the exponential distribution, allowing for heavy-tails.
For further information on the Mittag-Leffler distribution, see \citet{mittagleffler1}.
 
The most commonly used estimation methods for the Mittag-Leffler parameters so far are the method of log-moments, introduced by \citet{logmoment}, and maximum likelihood. 
The maximum likelihood estimator outperforms the method of log-moments in terms of mean squared error (MSE), but it is much more computationally intensive \citep{hees2020}. 
\citet{Cahoy2} showed that the log-moment estimator is asymptotically unbiased and asymptotically normal. 
Furthermore, the fractional moment estimator proposed by \citet{kozubowski} is also available; however, it has been shown to be outperformed by the log-moment estimator in terms of bias and MSE (see \cite{Cahoy2}). Therefore, we do not consider it here. 

\citet{christina} propose minimizing the Cramèr-von Mises distance $\Delta^{[\text{\tiny CM}]}$ for parameter estimation of the fractional compound Poisson process, which corresponds to a mixture distribution between the Dirac measure at zero and the Mittag-Leffler distribution. We investigate this estimation method in our pure Mittag-Leffler setting and include it in the simulation study presented in \mbox{Section \ref{sec:simulation_2}}. The Cramèr-von Mises (CM) estimator is given by 
\begin{align*}
 \left( \begin{array}{c}
\hat{\beta}_{\text{CM}}\\
\hat{\sigma}_{\text{CM}}\\
\end{array} \right)
& = \underset{ \left(\begin{smallmatrix}\beta\\\sigma\end{smallmatrix}\right) \in (0,1] \times \mathbb{R}^+}{\text{argmin}}
	\Delta^{[\text{\tiny CM}]}\left(F_{n} , F_{\beta,\sigma}\right)\\
    &= \underset{ \left(\begin{smallmatrix}\beta\\\sigma\end{smallmatrix}\right) \in (0,1] \times \mathbb{R}^+}{\text{argmin}} \sum_{i = 1}^n\left(\frac{2i-1}{2n} - F_{\beta,\sigma}\left(x_{(i)}\right) \right)^2,
\end{align*}
where $F_{\beta,\sigma}$ denotes the CDF of the Mittag-Leffler distribution with tail parameter $\beta$ and scale parameter $\sigma$, $F_n$ represents the empirical CDF, and $x_{(1)}, \ldots ,x_{(n)} $ are the $n$ ordered observations.

Quantiles are important characteristics of statistical distributions and provide important information e.g., on the location and the variability of its outcomes. As opposed to the ordinary moments of the distribution, they do always exist, and the sample quantiles are rather robust except for extreme ones. 
Therefore, we propose quantile based estimation of the Mittag-Leffler parameters.
Similar approaches have already been suggested in other settings. For example, \citet{Dubey} fits the Weibull distribution analytically by matching two quantiles of the Weibull distribution to the corresponding empirical quantiles.
Recently, \citet{QLS} elaborated quantile least squares (QLS) estimation for location-scale families, which numerically minimizes the sum of the squared distances between $r \geq 2$ empirical and theoretical quantiles.
This estimator is given by \[ 
 \left( \begin{array}{c}
\hat{\beta}_{\text{QLS}}\\
\hat{\sigma}_{\text{QLS}}\\
\end{array} \right) = \underset{ \left(\begin{smallmatrix}\beta\\\sigma\end{smallmatrix}\right) \in (0,1] \times \mathbb{R}^+}{\text{argmin}} \sum\limits_{i = 1}^r \bigl( F^{-1}_{\beta, \sigma}(\alpha_i) -\hat{q}_{\alpha_i}\bigr) ^2,\]
where $\hat{q}_{\alpha_1}, \ldots, \hat{q}_{\alpha_r}$ denote the empirical quantiles with $0 <\alpha_1<\ldots <\alpha_r  < 1$. We examine the performance of the QLS estimator in the Mittag-Leffler setting and include it in the simulation study presented in \mbox{Section \ref{sec:simulation_2}}.

We propose an estimator, which combines ideas of the CM and the QLS approach. It is based on numerical minimization of the sum of squared distances between the CDF of the model distribution and the empirical CDF at $r$ empirical quantiles $\hat{q}_{\alpha_1}, \ldots, \hat{q}_{\alpha_r}$.
This quantile-based (QB) estimator for the Mittag-Leffler parameters is defined as
 \begin{equation}
 \left( \begin{array}{c}
\hat{\beta}_{\text{QB}}\\
\hat{\sigma}_{\text{QB}}\\
\end{array} \right)
 = \underset{ \left(\begin{smallmatrix}\beta\\\sigma\end{smallmatrix}\right) \in (0,1] \times \mathbb{R}^+}{\text{argmin}} \sum\limits_{i = 1}^r \bigl( \alpha_i - F_{\beta,\sigma}(\hat{q}_{\alpha_i}) \bigr) ^2.\label{eq:QLS}\end{equation}
The QB estimator can be regarded as a simplification of the CM estimator, as it also compares the empirical and model CDFs, but focuses on some selected quantiles rather than considering all order statistics. 
The optimization is done with the L-BFGS-B algorithm introduced in \citet{L-BFGS-B}. In \mbox{Section 3}, we demonstrate that using more than two quantiles is more informative and reduces the empirical MSE of our parameter estimators. However, if the quantiles are selected appropriately, each additional increase in $r$ provides progressively less additional information. Based on our simulation studies, we found that using $r = 5$ quantiles in the QB estimator produces a rather efficient estimator for the Mittag-Leffler parameters while maintaining a low computation time. Given the many possible choices for $\alpha_1,\ldots,\alpha_r$, we further address this issue in \mbox{Section \ref{sec:simulation_1}}. 
In Section~\ref{consistency} of the appendix, we prove that the QB estimator is consistent and asymptotically normal, provided that the quantiles are chosen appropriately. In case of the Mittag-Leffler distribution, a sufficient condition for consistency is that, for at least one pair of quantiles $q_{\alpha_i}$ and $q_{\alpha_j}$ used,
\[\frac{\partial}{\partial \sigma } F_{\beta,\sigma}(q_{\alpha_i})  \frac{\partial}{\partial \beta } F_{\beta,\sigma}(q_{\alpha_j}) \neq  \frac{\partial}{\partial \sigma} F_{\beta,\sigma}(q_{\alpha_j})  \frac{\partial}{\partial \beta} F_{\beta,\sigma}(q_{\alpha_i})\]
holds. For asymptotic normality, we additionally require that each of the inequalities
\begin{align*}
    \frac{\partial}{\partial \sigma}F(q_{\alpha_j}) \sum_{i = 1}^r \left(\frac{\partial}{\partial \beta}F(q_{\alpha_i})\right)^2 \neq \frac{\partial}{\partial \beta}F(q_{\alpha_j}) \sum_{i = 1}^r \frac{\partial}{\partial \beta}F(q_{\alpha_i})\frac{\partial}{\partial \sigma}F(q_{\alpha_i}), \\
    \frac{\partial}{\partial \beta}F(q_{\alpha_j}) \sum_{i = 1}^r \left(\frac{\partial}{\partial \sigma}F(q_{\alpha_i})\right)^2 \neq \frac{\partial}{\partial \sigma}F(q_{\alpha_j}) \sum_{i = 1}^r \frac{\partial}{\partial \beta}F(q_{\alpha_i})\frac{\partial}{\partial \sigma}F(q_{\alpha_i})\phantom{,}
\end{align*}
is satisfied for at least one quantile $q_{\alpha_j}$. We conjecture that these conditions are met whenever $\alpha_1, \ldots, \alpha_r$ are chosen such that at least one $\alpha_i < 0.1797$ and one $\alpha_j > 0.5935$ (for details, see Sections~\ref{consistency} and \ref{sign_derivative} in the appendix). Due to the asymptotic normality, confidence intervals can be constructed under the conditions stated above. For further details on their construction and performance, see Section~\ref{consistency}.

\subsection{Fractional Poisson Process with Seasonal Variation}
\label{sec:methods_3}

The fractional Poisson process, first defined by \citet{fpp_original3}, offers a natural generalization of the widely used Poisson process. It enables the description of clustering behavior in the occurrences of events, often found in real-world applications.
Consider a stochastic process $(X_m)_{m \in \mathbb{N}}$ of real-valued events and let $(T_m)_{m \in \mathbb{N}}$ denote the sequence of increasing times at which these events occur. Then, the return times $(W_m)_{m \in \mathbb{N}}$ are defined as the times \mbox{$W_{m} = T_{m+1} - T_{m}$} between two consecutive 
events.
In the Poisson process, the return times $(W_m)_{m \in \mathbb{N}}$ are independent and follow an exponential distribution. 
The fractional Poisson process generalizes this to the Mittag-Leffler distribution, which contains the exponential distribution as a limiting case for $\beta=1$. The fractional Poisson process thus extends the Poisson process to heavy-tailed return times.
For more detailed information on the fractional Poisson process, we refer to \citet{fpp_original}, \citet{fpp} and \citet{fpp2}.

In many real-world applications, the observed process exhibits seasonal fluctuations, especially in meteorological contexts.
To effectively address these scenarios, we modify the fractional Poisson process by allowing the parameters $\beta$ and $\sigma$ of the Mittag-Leffler distribution to vary periodically over time instead of remaining constant.
To avoid restrictive assumptions about the nature of the underlying seasonality, we estimate the parameters using a flexible nonparametric approach and allow each day of the calendar year to have its own parameter tuple,
\[\left(\begin{smallmatrix}
\beta(1)\\\sigma(1)\end{smallmatrix}\right), \ldots , \left(\begin{smallmatrix}\beta(365)\\\sigma(365)
\end{smallmatrix} \right),\]
where the number indicates the day of the calendar year, starting January 1.
To estimate the parameters $\beta(t_0)$ and $\sigma(t_0)$ on day $t_0$, we consider the return times starting in a time window centered at $t_0$.
Return times starting at a time close to $t_0$ should have a stronger influence on the estimate $(\hat{\beta}(t_0),\hat{\sigma}(t_0))$ than those farther away. 
We use an Epanechnikov kernel \citep{epanechnikov} to assign weights based on the distance between $t$ and $t_0$.

More precisely, let $t_1, \ldots, t_n \in \{ 1, \ldots , 365 \}$ be the calendar days at which the events $x_1,\ldots, x_n$ with the corresponding return times $w_1,\ldots, w_n$ have been observed. Since the observation times are random, it is possible to have one, multiple, or no observations on a given day $t_0$.
Due to the cyclical nature of the calendar year, the distance between two days $h,j \in \{1, \ldots , 365 \}$ is given by 
\[d(h,j) = \min(|h-j|, 365 - |h-j|  ).\]
Then the weight of the $i$-th observation used to calculate the estimates on day $j$ is given by
\[k_{c}(j, t_i) = \frac{3}{4c}\cdot \max\left(0, 1-\left(\frac{d(t_i,j)}{c}\right)^2\right),\]
where $c > 0$ is the bandwidth of the Epanechnikov kernel.
Only observations made less than $c$ days apart from $j$ influence $\hat{\beta}(j)$ and $\hat{\sigma}(j)$.
The weights $k_{c}(j, t_1), \ldots, k_{c}(j, t_n)$ are normalized for each day separately, and are given by
\[ \bar{k}_{c}(j, t_i) = \frac{k_{c}(j, t_i)}{\sum\limits_{l = 1}^n k_{c}(j, t_l)} \quad \text{for } j = 1,\ldots , 365.\]
For each calendar day of the year, the normalized weights $ \bar{k}_{c}(j,t_1), \ldots,  \bar{k}_{c}(j,t_n)$, which depend on the observations times $t_1, \ldots, t_n$, are assigned to the return times $w_1,\ldots, w_n$ and used in the weighted parameter estimation. We construct kernel weighted versions of the log-moment, the maximum likelihood, the quantile least squares, and the quantile-based estimator using the following modifications.

The log-moment estimator is adapted by replacing the first and second moments of the logarithmized observations with their respective weighted versions, defined as
\[\hat{\mu}_j = \sum\limits_{i=1}^n \bar{k}_{c}(j,t_i)\cdot \ln(w_i) \quad \text{and} \quad \hat{\sigma}^2_j = \frac{1}{1-V} \sum\limits_{i=1}^n \bar{k}_{c}(j,t_i) \cdot (\ln(w_i)-\hat{\mu}_j)^2,\]
where $V = \sum \limits_{i=1}^n \bar{k}_{c}(j,t_i)^2$. The weighted variance is determined as implemented in \citet{mitey}. Otherwise, the estimation procedure remains as described in ~\citet{logmoment}, applied for each day $j = 1, \ldots, 365$.

Weights can be included in the maximum likelihood estimator by calculating
\[ \left(\begin{smallmatrix}\hat{\beta}(j)\\
\hat{\sigma}(j)
\end{smallmatrix} \right) =\underset{  \left(\begin{smallmatrix}\beta\\\sigma\end{smallmatrix}\right) \in (0,1] \times \mathbb{R}^+}{\text{argmax}} \sum_{i = 1}^n \bar{k}_{c}(j,t_i)\cdot \ln \left(f_{\beta, \sigma}(w_i) \right).\]

For the quantile least squares estimator, we use weighted quantiles similarly to the weighted median used, e.g., in \citet{weighted_median}. Let $w_{(1)} \leq \ldots \leq w_{(n)}$ denote the ordered return times and let $\bar{k}_c(j, t_{(1)}), \ldots, \bar{k}_c(j, t_{(n)})$ denote the corresponding weights on day $j$. Then, the weighted empirical $\alpha$-quantile $\hat{q}_{\alpha}(j)$ with weights $\bar{k}_c(j, t_{(1)}), \ldots, \bar{k}_c(j, t_{(n)})$ is defined as 
\[ \hat{q}_{\alpha}(j) = w_{(\ell)} \text{ with } \ell = \max \{h : \sum \limits_{i = h}^n \bar{k}_c(j, t_{(i)}) \geq 1-\alpha \}.\]
The weighted empirical quantiles replace the ordinary empirical quantiles in the quantile-based approach described in (\ref{eq:QLS}) for the daily parameter estimation. We call the arising method weighted quantile-based estimator.

When both seasonality and trend are taken into account, this approach can be extended in a similar manner. However, it should be noted that the Epanechnikov kernel should not be applied within $c$ days of the beginning or end of the observation period. As an alternative, weighting based on a beta kernel can be employed at the boundaries of the observation period \citep{beta_kernel}. 
	\section{Simulation Study}\label{sec:simulation}

In this section, we first examine the selection of quantiles for the QB estimator in a simulation study. Afterward, we compare the performances of the log-moment (LM), maximum likelihood (ML), Cramèr-von Mises (CM), quantile least squares (QLS) and quantile-based (QB) estimators. The R ~\citep{R} package \texttt{MittagLeffleR} ~\citep{MittagLeffleR} was used to generate observations from a Mittag-Leffler distribution and to compute the log-moment and maximum likelihood estimators. The R code used in this simulation study is available via the link provided in the Data Availability Statement.

\subsection{Settings}\label{sec:settings}
We conduct a simulation study to evaluate different quantile combinations in the quantile-based estimator and to compare the estimators for the Mittag-Leffler parameters. 
We consider Mittag-Leffler distributions with different tail parameters $\beta$ and scale parameters $\sigma$ and several sample sizes $n$.
Specifically, we consider all combinations of
\begin{itemize}
    \item $\beta \in A_{\beta} = \{0.6, 0.65,  \ldots, 1 \} $,
    \item $\sigma \in  A_{\sigma} = \{25, 50, 100, 250, 500, 750, 1000, 1500, 2000 \} $ and 
    \item $n \in  A_{n} = \{200, 500, 1000, 5000\} $,
\end{itemize}
resulting in a total of $| A_{\beta}| \cdot | A_{\sigma}| \cdot | A_{n}| =  9 \cdot 9 \cdot 4 = 324 $ different settings.
These settings cover parameter values of $\beta$ and $\sigma$ that, based on preliminary examinations, we assess as realistic in the context of relative vorticity extremes. For each setting, we generate a total of 1000 datasets.

\subsection{Selecting Quantiles for the QB Estimator} \label{sec:simulation_1}

First, we aim to identify suitable quantile combinations for the QB estimator. \citet{Dubey} derived the optimal pair of quantiles for parameter estimation in the case of Weibull-distributed variables analytically by minimizing the determinant of the asymptotic covariance matrix of the estimator.
However, this analytical approach is considerably more complicated in the case of the Mittag-Leffler distribution. 
Hence, we estimate the MSEs of the estimators arising from different quantile combinations numerically for various scenarios instead. For each setting, the MSE is estimated for both parameters using 1000 datasets generated for the corresponding setting, resulting in 324 empirical MSEs per estimator for both the tail and the scale. We aim to identify quantile combinations that yield low MSEs across the considered settings for both parameters. 

\begin{figure}[ht]
    \centering
    \includegraphics[width = 0.9\linewidth]{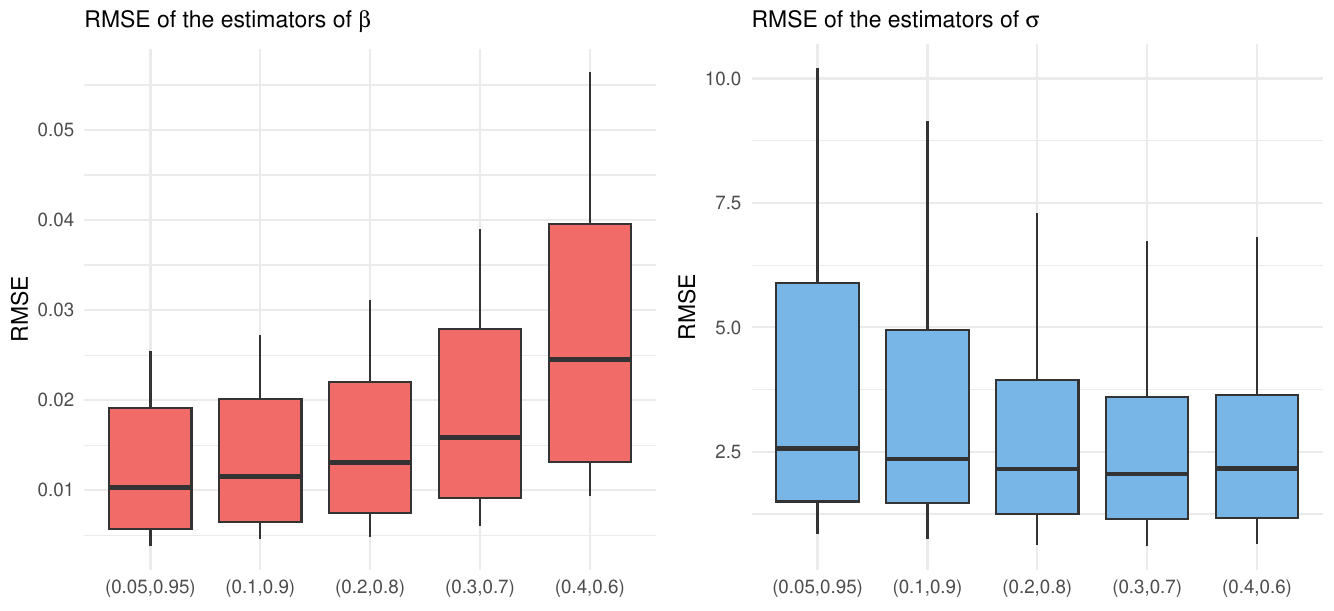}
    \caption{Empirical RMSE of five QB estimators for the tail (left) and the scale (right) parameter, all estimators utilizing two quantiles $\left(\hat{q}_{\alpha_1}, \hat{q}_{\alpha_2} \right)$.}
    \label{fig_exm}
\end{figure}

First we consider pairs of quantiles, i.e., $r=2$. Our analysis indicates the intuitive result that a combination of a high and a low quantile results in lower MSEs for the tail parameter, whereas using more central quantiles is advantageous for estimation of the scale parameter. 
This is illustrated in \mbox{Figure \ref{fig_exm}}, which shows boxplots of the empirical root mean squared errors (RMSEs) 
for the tail (left) and the scale parameter (right)
of five QB estimators corresponding to the quantile combinations  
\[(\alpha_1, \alpha_2) \in \{(0.05,0.95) , (0.1,0.9), ( 0.2, 0.8), (0.3,0.7),( 0.4,  0.6) \}.\] 
\mbox{Figure \ref{fig_exm}} is restricted to settings with $\sigma = 50$, since including all settings would complicate the visual comparison of the empirical RMSEs of the scale estimators due to the broad variety of values considered for the scale parameter. This reduces the number of settings considered to 36. Obviously, the quantile combinations $(0.05,0.95)$ and $(0.1, 0.9)$ yield the lowest empirical RMSEs for the tail parameter.
In contrast, the quantile combinations $( 0.3,  0.7)$ and $(0.4, 0.6)$ outperform the others for the scale, while the combination $(0.2, 0.8)$ might serve as a compromise.

In view of these findings, we investigate whether the parameter estimation can be improved further by combining low, central, and high quantiles.
Our results indicate that using five quantiles enhances simultaneous parameter estimation while increasing the computation time only slightly. We build five sets of low, low to central, central, central to high, and high quantiles and consider all 3125 combinations of $\alpha_1, \ldots, \alpha_5$ from the following sets:
\begin{itemize}
    \item $\alpha_1 \in \{0.05,\, 0.075,\, 0.1,\, 0.125,\, 0.15 \} $,
    \item $\alpha_2 \in \{0.2, \,0.225, \,0.25, \,0.275,\, 0.3 \} $,
    \item $\alpha_3 \in \{0.45,\, 0.475,\, 0.5,\, 0.525, \,0.55 \} $,
    \item $\alpha_4 \in \{0.7, \,0.725,\, 0.75,\, 0.775,\, 0.8 \} $,
    \item $\alpha_5 \in \{0.85, \,0.875,\, 0.9,\, 0.925,\, 0.95 \} $.
\end{itemize}  
We assess the performance of the different quantile combinations based on the empirical efficiency of the QB estimator relative to the ML estimator, which performs best in terms of MSE in the Mittag-Leffler setting ~\citep{hees2020}.
The tail and the scale parameter should have the same influence on the evaluation. We thus add the empirical relative efficiencies of $\hat{\beta}_\text{QB}$ and $\hat{\sigma}_\text{QB}$ compared with the ML estimator and take the average over all settings, \begin{equation}
\frac{1}{324} \sum_{i = 1}^{324} \left(\frac{\widehat{\text{MSE}}(\hat{\beta}_{\text{ML}_i})}{\widehat{\text{MSE}}(\hat{\beta}_{\text{QB}_i})} +\frac{\widehat{\text{MSE}}(\hat{\sigma}_{\text{ML}_i})}{\widehat{\text{MSE}}(\hat{\sigma}_{\text{QB}_i})}\right), \label{comp} \end{equation}
where $\hat{\beta}_{\text{ML}_i}$ and $\hat{\sigma}_{\text{ML}_i}$ denote the ML and $\hat{\beta}_{\text{QB}_i}$ and $\hat{\sigma}_{\text{QB}_i}$ the QB estimators in the $i$-th setting.
Alternatively, a measure that accounts for the dependence between $\hat{\beta}$ and $\hat{\sigma}$ is the determinant of their covariance matrix
\begin{equation*}\text{Var}(\hat{\beta})\text{Var}(\hat{\sigma}) - \text{Cov}(\hat{\beta}, \hat{\sigma}) ^2.  \label{extra1} 
\end{equation*}
This is considered, for example, in the evaluation of the percentile estimators for the Weibull parameters in \citet{Dubey}. However, since this measure does not account for a potential bias, we instead consider the determinant of the mean squared error matrix
\begin{equation}\text{det}\begin{pmatrix}
\text{E}[(\hat{\beta} - \beta)^2]  & \text{E}[(\hat{\beta} - \beta)(\hat{\sigma} - \sigma)]  \\
\text{E}[(\hat{\beta} - \beta)(\hat{\sigma} - \sigma)] & \text{E}[(\hat{\sigma} - \sigma)^2] \end{pmatrix},\tag{$\star\star$} \label{extra2}
\end{equation}which reduces to the determinant of the covariance matrix when the estimators are unbiased. Note that the diagonal elements correspond to the MSEs of the estimators, while the off-diagonal elements represent the second-order cross-moment of the estimation errors, which coincides with the covariance in the unbiased case.
We examine the average relative performance of criterion (\ref{extra2}) of the ML and the QB estimators, measured by the average ratio
\begin{equation}
\frac{1}{324} \sum_{i = 1}^{324} \left(\frac{\widehat{\text{MSE}}(\hat{\beta}_{\text{ML}_i})\widehat{\text{MSE}}(\hat{\sigma}_{\text{ML}_i}) - \widehat{\text{E}}[(\hat{\beta}_{\text{ML}_i} - \beta)(\hat{\sigma}_{\text{ML}_i} - \sigma)]^2}{\widehat{\text{MSE}}(\hat{\beta}_{\text{QB}_i})\widehat{\text{MSE}}(\hat{\sigma}_{\text{QB}_i}) - \widehat{\text{E}}[(\hat{\beta}_{\text{QB}_i} - \beta)(\hat{\sigma}_{\text{QB}_i} - \sigma)]^2}\right). \label{extra} 
\end{equation}
Among the 3125 quantile combinations considered, criterion (\ref{comp}) is maximized by 
\[(\alpha_1,\, \alpha_2,\, \alpha_3,\, \alpha_4 ,\, \alpha_5 ) = (0.1, \,0.3, \,0.5,\, 0.8,\, 0.925) ,\]
which has the eighth-best performance in terms of criterion (\ref{extra}).
Other combinations that perform particularly well regarding the metrics in (\ref{comp}) and (\ref{extra}) are \[(0.1, \,0.3, \,0.525,\, 0.8,\, 0.925), \, (0.05, \,0.2, \,0.475,\, 0.8,\, 0.925) \text{ and } (0.05, \,0.2, \,0.5,\, 0.8,\, 0.925).\]
It is not surprising that the best-performing combinations are quite similar to each other.
The 25 quantile combinations that yield the best results in terms of metric (\ref{comp}) all use $\alpha_4 = 0.8$ and $\alpha_5 = 0.925$, and none of them include $\alpha_1 = 0.15$, see Table~\ref{25comb} in the appendix (page~\pageref{25comb}). The combinations in Table~\ref{25comb} generally also perform well with respect to criterion~(\ref{extra}). Consequently, including the dependence between $\hat{\beta}$ and $\hat{\sigma}$ has only a minor effect on the assessment of the quantile combinations.

In general, we recommend combining high, central, and low quantiles for the QB estimation of both Mittag-Leffler parameters. In the following sections, we use the $(0.1, \,0.3, \,0.5,\, 0.8,\, 0.925)$ quantiles due to the good performance in terms of the average empirical efficiency relative to the ML estimator.

\subsection{Comparison of the Mittag-Leffler Estimation Methods} \label{sec:simulation_2}

Now we assess the performance of the CM, LM, ML, QLS and QB estimators in terms of MSE, computation time, and robustness. 
For the QLS estimator, we
follow the recommendation of \citet{QLS} to select quantiles equidistantly and use the following quantile combinations:
\[(0.1, \,0.3,\, 0.5, \,0.7, \,0.9 ) \text{ and } (0.05, \,0.275,\, 0.5,\, 0.725, \,0.95).\]
Additionally, we consider the same quantile combination as for the QB.

\begin{figure}[ht]
    \centering
    \includegraphics[width=0.92\linewidth]{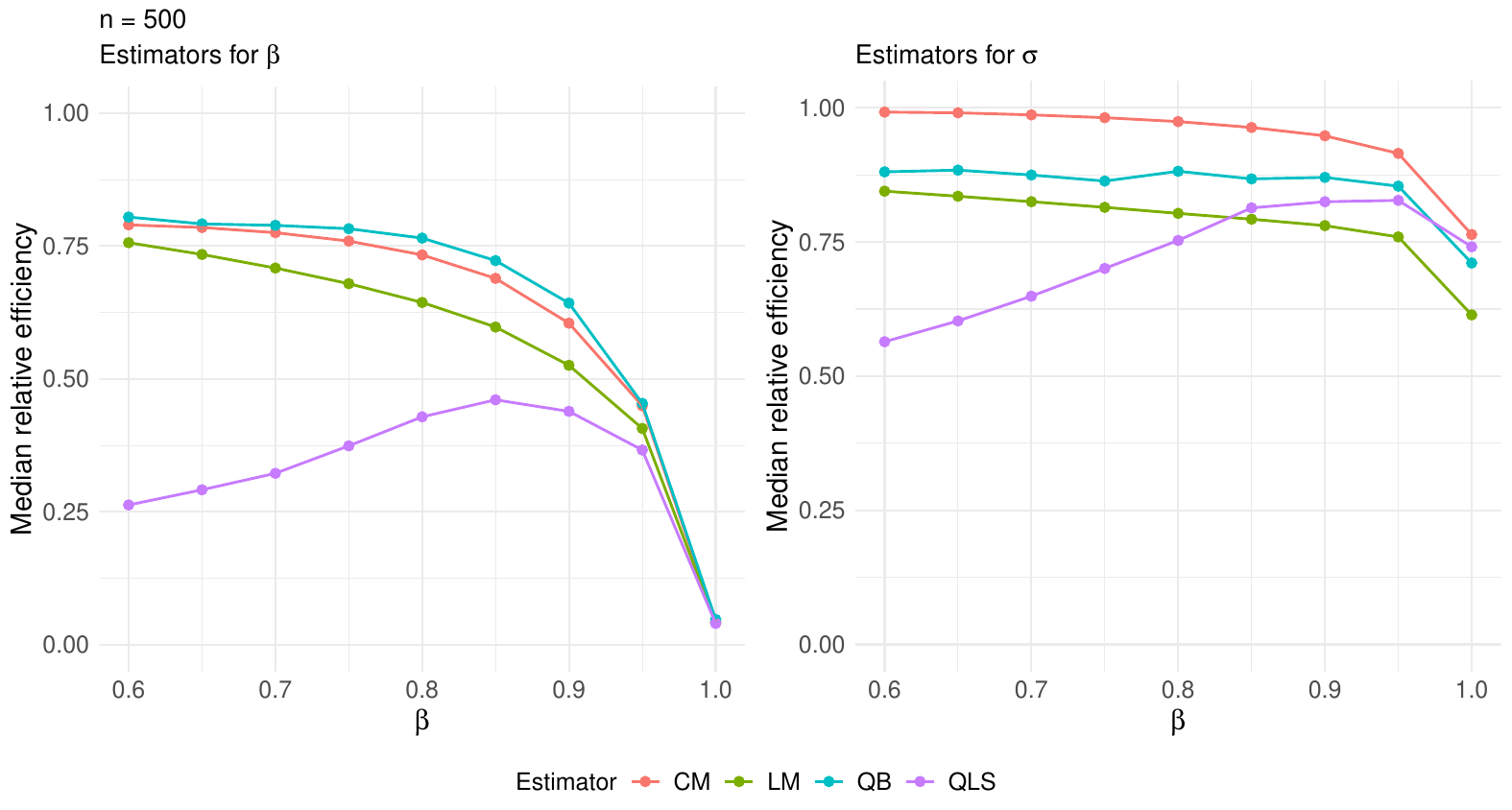}
    \caption{Median empirical relative efficiency of the CM, LM, QB and QLS estimators for the tail parameter $\beta$ (left) and the scale parameter $\sigma$ (right), evaluated across different values of the underlying tail parameter $\beta$, with $n = 500$.}
    \label{fig_efficiency1}
\end{figure}

We again use the settings described in \mbox{Section \ref{sec:settings}} and obtain 324 empirical MSEs per estimator for both parameters.
We calculate the empirical relative efficiency of each estimator with respect to the ML estimator in every setting given by
\[\widehat{\text{e}}(\hat{\beta}_i) = \frac{\widehat{\text{MSE}}(\hat{\beta}_{\text{ML}_i})}{\widehat{\text{MSE}}(\hat{\beta}_i)} \quad \text{and} \quad \widehat{\text{e}}(\hat{\sigma}_i)=\frac{\widehat{\text{MSE}}(\hat{\sigma}_{\text{ML}_i})}{\widehat{\text{MSE}}(\hat{\sigma}_i)}.\]Figure~\ref{fig_efficiency1} depicts median relative efficiencies of the CM, LM, QB and QLS estimators for the tail parameter $\beta$ (left) and the scale parameter $\sigma$ (right) 
as a function of the true underlying tail parameter value, considering the 81 settings where $n = 500$. For the QLS estimator, only the results for the $(0.1, 0.3, 0.5, 0.7, 0.9)$ quantile combination are given, as this combination showed the best performance among those considered.

For the tail parameter, the CM and QB estimators exhibit similar median relative efficiency across all values of $\beta$, with the QB estimator performing slightly better overall. Both consistently outperform the LM and QLS estimators for $\beta \neq 1$.
For the CM, QB and LM estimators, the median relative efficiency decreases with increasing $\beta$.
The QLS estimator performs notably worse in the tail parameter estimation, especially for smaller values of $\beta$, achieving a maximum median relative efficiency of only 46.1\% at $\beta = 0.85$.
In the special case of $\beta = 1$, i.e., the exponential distribution, the median relative efficiency of all estimators for the tail parameter is below 5\%. 
This is due to a substantial improvement in the MSE of the ML estimator in this special case, rather than a decline in the performance of the other estimators.
Moreover, the relative efficiency of all estimators is nearly constant for the same value of the tail parameter, with the largest observed range being 1.35\%, so that the median represents the overall relative efficiency quite accurately. 

For the scale parameter, the CM estimator achieves the highest median relative efficiency for all values of $\beta$, exceeding 95\% for $\beta \leq 0.85$. 
In case of heavy-tailed distributions, where $\beta < 1$, the QB estimator shows the second-best performance, with median relative efficiencies above 85\%.
For small values of the tail parameter ($\beta \leq 0.8$), the QLS estimator has the lowest median efficiency, while for $\beta \geq 0.85$, the LM estimator performs worst in this respect.
As with the tail parameter, the median relative efficiency of all estimators drops at $\beta = 1$, but it remains above 60\% for all estimators. 
As for the tail parameter, the relative efficiencies of the CM, LM and QLS estimators are almost identical for distributions with the same tail parameter, with the largest observed range being 1.62\%. In contrast, the relative efficiencies of the QB estimator exhibit a maximum range of 17.36\% for the same value of $\beta$. The QLS estimator exhibits a slightly higher relative efficiency
than the QB estimator only in the exponential case $\beta=1$, while the QB outperforms the QLS for both parameters whenever the underlying distribution is heavy-tailed. 
This suggests that, for the Mittag-Leffler distribution, when heavy tails are present, minimizing the distance between the empirical and the theoretical CDF yields better results, whereas in the absence of heavy tails, comparing empirical and theoretical quantiles is more effective.

The results for the settings with $n = 200,\, 1000$ and $5000$ are displayed in Figures~\ref{fig_efficiency2}-\ref{fig_efficiency4} in the appendix (pages~\pageref{fig_efficiency2}-\pageref{fig_efficiency4}). For both parameters, the curves are quite similar to those for $n= 500$, presented in Figure~\ref{fig_efficiency1}. As before, the QLS estimator is clearly outperformed by the other estimators for the tail parameter, with maximum median relative efficiencies of 53.2\% ($n = 200$), 48.1\% ($n = 1000$) and 49.0\% ($n = 5000$). The CM, QB and LM estimators again have higher median relative efficiencies for smaller values of the tail parameter, and the CM and the QB consistently outperform the LM estimator. All these estimators exhibit very low relative efficiencies in the special case of $\beta = 1$, which, as before, is due to a substantial decrease in the MSE of the ML estimator, rather than an increase in the MSE of the others. 
In the estimation of the scale parameter, the CM estimator again performs similarly well to the ML estimator, with median relative efficiency of approximately 100\% in the settings with $n = 1000, \,5000$ and $\beta = 0.6,\, 0.65, \,0.7$. 
The QB estimator shows the second-best performance in terms of median relative efficiencies for most values of the tail parameter, albeit the QLS estimator surpasses it if $\beta$ is close to 1. As in the case of $n = 500$, the LM is outperformed by both the CM and the QB estimators in terms of median relative efficiency across all values of $\beta$ and sample sizes. Moreover, for larger values of $\beta$, the LM estimator exhibits the poorest performance in estimating the scale parameter.

Additionally, Figure~\ref{fig_res_metric3} in the appendix (page~\pageref{fig_res_metric3}) shows the performance of the estimators under criterion~(\ref{extra}), which jointly measures the relative efficiency of $\hat{\beta}$ and $\hat{\sigma}$ while accounting for their dependence. Under this criterion, the QLS estimator again performs worst for most values of $\beta$. The CM and QB estimators outperform the LM estimator across all sample sizes and values of $\beta$. For smaller values of $\beta$, the CM yields better results than the QB estimator regarding criterion~(\ref{extra}), however, for $\beta \geq 0.85$, their performance is similar.

\begin{figure}[ht]
    \centering
    \includegraphics[width=0.9\linewidth]{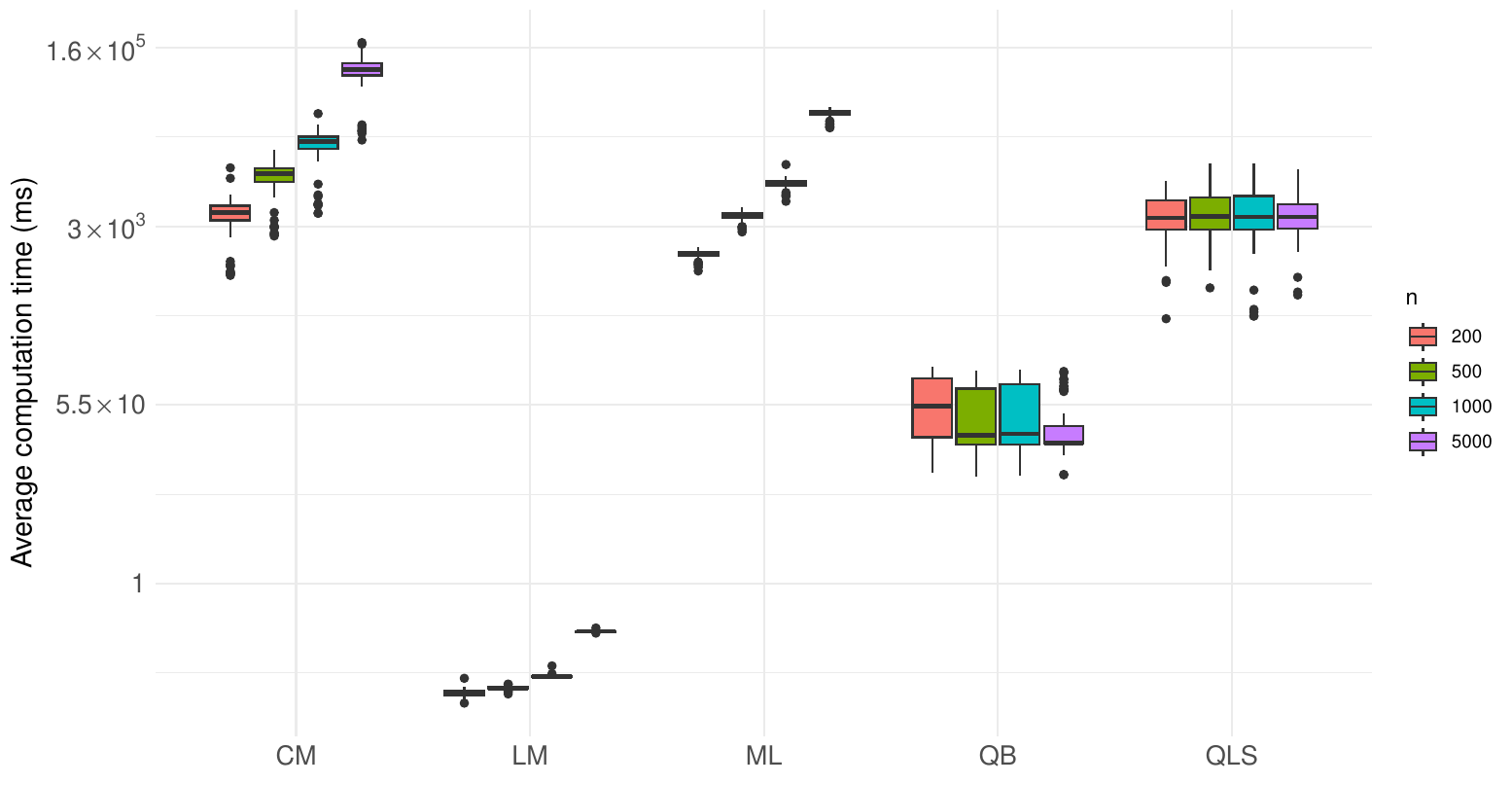}
    \caption{Average computation time in milliseconds (ms) of the CM, LM, ML, QB and QLS estimators for different sample sizes on a logarithmic scale.}
    \label{comp_time}
\end{figure}

Overall, the QB and the CM estimator perform better than the commonly used LM estimator with respect to criterion~(\ref{extra}) as well as the empirical relative efficiency for each parameter separately in the simulation study. In contrast, the QLS estimator performs poorly, particularly in the estimation of the tail parameter. Only the CM estimator achieves empirical relative efficiencies close to 100\% in individual settings, indicating that the ML estimator generally yields the lowest MSE among all estimators considered. However, the computation time of the ML estimator is substantially higher than that of the LM and QB estimators. This is illustrated in Figure \ref{comp_time}, where boxplots of the average computation times are shown for different sample sizes. Each boxplot is based on 81 average computation times, corresponding to the 81 settings for the given sample size. 

The CM estimator exhibits the largest average computation times for all sample sizes, reaching values of up to 182.60~s (setting with $\beta =1$, $\sigma = 1500$ and $n = 5000$).
For $n = 1000$ and $n=5000$, the ML estimator has the second-highest computation time. For both the CM and the ML estimator, the computation time increases substantially with the sample size. 
In contrast, the computation times of the QB and the QLS do not increase with the sample size in the settings considered. This is because the optimization requires fewer iterations for larger sample sizes, which compensates for the longer time needed to determine the sample quantiles.
Interestingly, computing the QLS estimator takes, on average, 133 times longer than computing the QB estimator, despite both using the same number of quantiles in the optimization.
The LM estimator requires the shortest computation time, which increases with the sample size. Since the other four estimators use the LM estimate as the initial value in their optimization, their computation times necessarily exceed that of the LM estimator.

\begin{figure}[ht]
    \centering
    \includegraphics[width=0.92\linewidth]{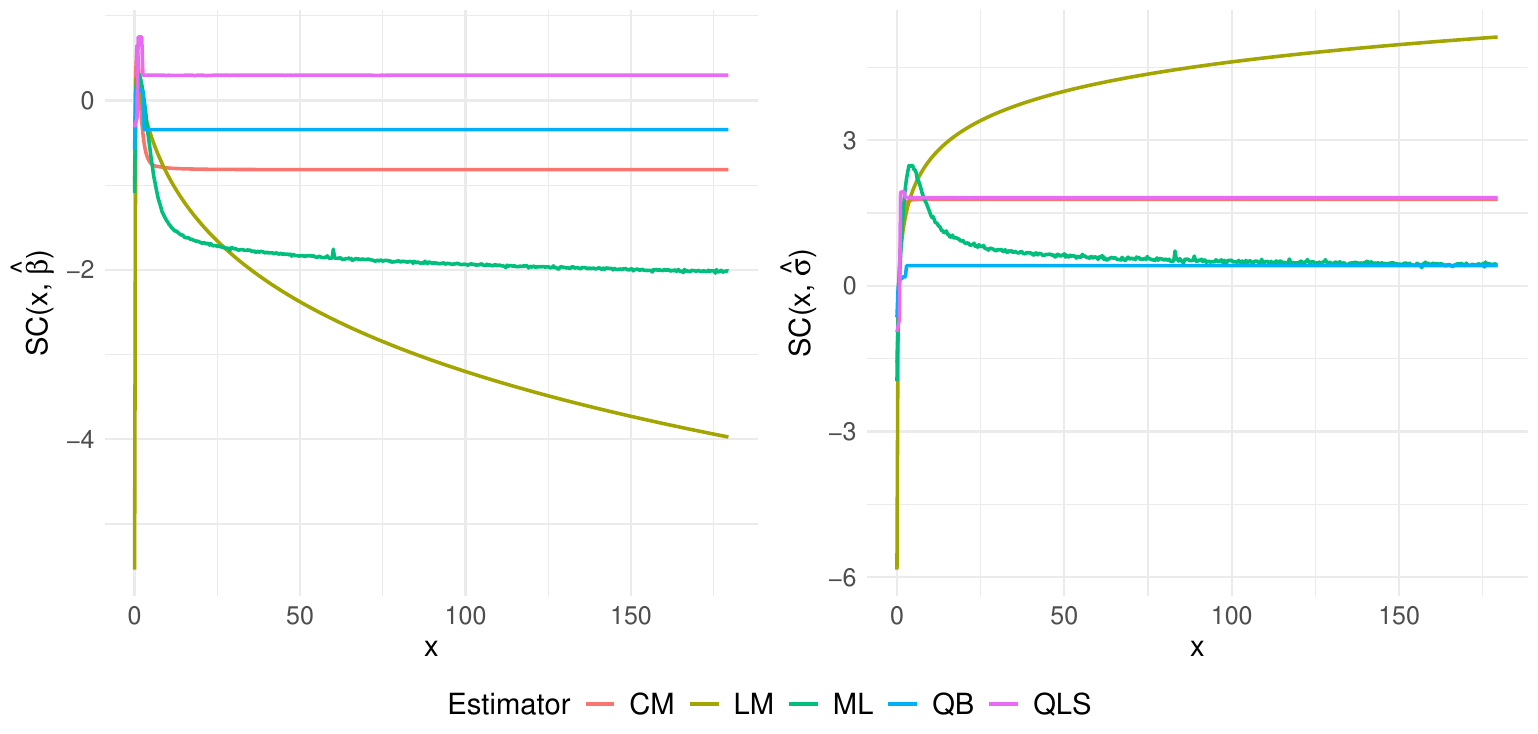}
    \caption{Sensitivity curves of $\hat{\beta}$ (left) and $\hat{\sigma}$ (right) of the CM, LM, ML, QB and QLS estimators.}
    \label{SC_1108}
\end{figure}

To evaluate the robustness of the different estimators, we consider the sensitivity curve (SC). The SC of an estimator $\hat{\beta}$ is given by
\begin{equation}\text{SC}\left(x, \hat{\beta} \right) = (n+1) \left(\hat{\beta}(x_1,\ldots, x_n, x) -\hat{\beta}(x_1,\ldots, x_n)\right),\label{SC} \end{equation}
where $ x_1,\ldots, x_n$ denote $n$ observations drawn from a Mittag-Leffler distribution. $\text{SC}(x, \hat{\sigma})$ is computed analogously.
To create sensitivity curves, we generate $n = 200$ observations from a Mittag-Leffler distribution with parameters $\beta = 0.9$ and $\sigma = 1$. The contamination point $x$ is varied over the range between the 0.001 and 0.999 quantiles of the distribution.
The resulting sensitivity curves for the different estimators are displayed in Figure~\ref{SC_1108}. 

For both parameters, the LM estimator is strongly influenced by the contamination point, with the sensitivity curves decreasing for one parameter and increasing for the other as $x$ increases. In contrast, the QB, QLS and CM estimators remain constant once $x$ exceeds a certain threshold. The ML estimator is less affected by $x$ than the LM estimator, but $\text{SC}(x, \hat{\beta}_\text{ML})$ still changes gradually as $x$ increases.

Overall, the CM estimator demonstrates better performance than the LM, QB and QLS estimators concerning the relative efficiency, but at the cost of a much higher computation time. 
The QB estimator can be viewed as a simplification of the CM estimator, as it uses much fewer points in the optimization. 
This implies a large reduction of the computation time, up to a factor of 8,000 in the scenarios considered here. Interestingly, we found the relative efficiency of both estimators to be similar for the tail parameter. There is some loss of efficiency when using the QB instead of the CM estimator for the scale parameter, but in the considered scenarios this loss is rather small. 
Both the QB and the CM estimators achieve consistently higher relative efficiencies than the commonly used LM estimator, while also showing robustness in the sensitivity curves.
The computation time of the ML estimator is up to 2,500 times higher than that of the QB estimator for the sample sizes considered here. Although both the ML and the CM outperform the QB estimator in terms of MSE, the latter can be recommended when computational resources are limited or when fitting Mittag-Leffler distributions to many large datasets, as in Section~\ref{sec:example}.

	\section{Analysis of Relative Vorticity Extremes}\label{sec:example}

Extratropical cyclones often exhibit clustering behavior, characterized by multiple cyclones passing over a given location within a short period of time. The temporal clustering of cyclones can lead to high precipitation totals and wind damage, affecting society both financially and in terms of human safety \citep{pinto1}.
In previous research, \citet{intro6} used the Poisson process to model occurrences of extratropical cyclones and found more serial clustering than expected by chance in the exit region of the North Atlantic storm track, Europe, and the central North Pacific.
Building on this, ~\citet{blender} proposed using the FPP to model boreal and austral winter vorticity extremes, which explicitly accounts for clustering behavior. 
This is closely connected to modeling extratropical cyclones, since in meteorology a cyclone in the Northern Hemisphere is typically identified by a maximum in the relative vorticity or a minimum in the mean sea-level pressure in a given area at a certain time (see e.g., \cite{Neu2013}). \citet{christina} applied an even more general model, the fractional compound Poisson process, to describe winter vorticity extremes, and used bootstrap methods to classify the regions into different submodels. Here, we focus on a region over the North Atlantic where they identified the FPP as the most appropriate model for the boreal winter vorticity extremes. However, severe storms are by no means limited to winter. Notable examples include the Great Storm of 15-16 October 1987 over southern UK \citep{great_storm_87}, Storm Christian on 25–30 October 2013 over northwestern Europe \citep{christian_2013}, and, more recently, Storm Ellen on 19–20 August 2020 over the UK and Ireland \citep{ellen_2020}.
Hence, we extend this research to a year-round examination of maxima in the relative vorticity in the North Atlantic-European region by incorporating seasonality into the FPP, as explained in Section~\ref{sec:methods_3}.

\subsection{Data and Method}
The following analysis is based on the relative vorticity at the \mbox{850 hPa} pressure level of the ERA5 reanalysis data \citep{datensatz} provided by the European Centre for Medium-Range Weather Forecast (ECMWF). We examine a part of the North Atlantic-European region (\mbox{45°~N\,–\,60°~N,} \mbox{40°~W\,–\,0°}) on a \mbox{1°\,$\times$\,1°} grid from January 1, 1940 to March 31, 2023. Instead of hourly vorticity values, we only include the observations made at \mbox{12 a.m.}, \mbox{6 a.m.}, \mbox{12 p.m.} and \mbox{6 p.m.} to reduce dependence in the data. 
The vorticity extremes are determined with the peaks over threshold method. As in \citet{blender} and ~\citet{christina}, all observations exceeding the $99\%$ quantile, calculated separately for each grid point, are classified as vorticity extremes, resulting in 1216 extremes per grid point. The ERA5 reanalysis data and the R code used in this study are available via the links provided in the Data Availability Statement.

We compute daily parameter estimates for each coordinate point separately using weights derived from an Epanechnikov kernel as outlined in \mbox{Section \ref{sec:methods_3}}. Choosing a larger bandwidth for the kernel, thereby including more days in the parameter estimation, increases bias, whereas using a smaller bandwidth increases variance. In the context of extratropical cyclones, it is standard to combine the data of a meteorological season, i.e., three months; see, e.g., \citet{blender}. To achieve comparability, we include 45 days before and 45 days after the day of interest in the estimation. Since the weight given to an observation decreases quadratically with the distance, observations at the edge of the moving window have much less influence on the estimates than observations in the center. 

To evaluate whether the seasonal pattern is stable throughout the observation period, we split the data into the periods from 1940–1981 (D1) and from 1982–2023 (D2). We compute estimates of the seasonal FPP for both periods separately and calculate the distances
 \begin{equation}
 \sum_{i= 1}^{365} \left(\hat{\beta}(t_i; \text{D1})-\hat{\beta}(t_i; \text{D2})\right)^2 \quad \text{and} \quad \sum_{i= 1}^{365} \left(\hat{\sigma}(t_i; \text{D1})-\hat{\sigma}(t_i; \text{D2})\right)^2.
 \label{permutation}\end{equation}
In order to assess whether these distances can occur just by chance, we conduct separate permutation tests for both parameters at each grid point. 
For this purpose, we randomly reorder the 84 years included in our analysis, so that a random selection of 42 years becomes the new period D1 and the other 42 years become D2. We calculate the distance (\ref{permutation}) for the reordered data and repeat this a total of 1,000 times for each grid point. Under the hypothesis of a stable seasonal pattern, we expect the estimates arising from all constructed subsets to be similar as the ordering of the years is just by chance. Our reordering causes some distortions at the beginning of the years but we expect these effects to be small.  
If the distance for the original data is unusually large as compared to the distances for the reordered data, i.e., if it falls within the 5\% largest distances obtained from the permuted datasets, the test rejects the null hypothesis of a constant seasonal pattern.

Due to the high computational cost of the 1,000 permutations for each grid point, we use the weighted QB estimator for the permutation tests.
In total, the permutation test rejects the null hypothesis for 5.03\% (tail parameter) and 8.99\% (scale parameter) of the grid points, which is slightly more than expected under the null hypothesis. However, the observations and thus the distances at adjacent coordinate points are not independent, so that the frequency of rejections might be somewhat increased due to larger variability. Therefore, we decide to consider a constant seasonal component in our analysis.

\subsection{Results}

In the following, we analyze the parameter estimates over the considered region on a large scale for four example days and discuss their underlying spatial structure. We then examine the annual course of the parameter estimates, with a more detailed focus given to a selected location. In addition to the estimates themselves, we also consider derived quantities, such as the probability of further vorticity extremes occurring within the next 72 hours. 

\begin{figure}[!htbp]
    \centering
    \includegraphics[width = \linewidth]{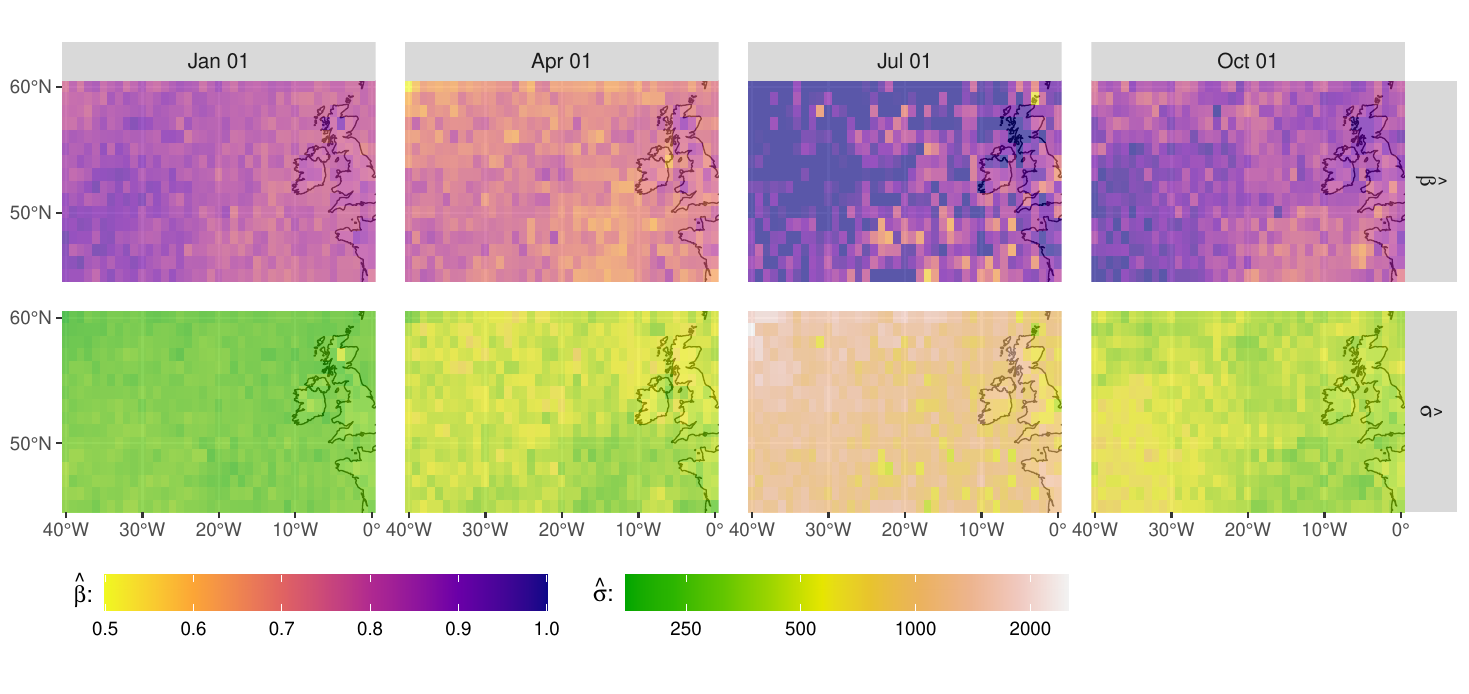}
    \caption{Heatmaps of the tail parameter (top) and scale parameter estimates (bottom) on January~1, April~1, July~1 and October~1 (from left to right).}
    \label{fig_map_eu}
\end{figure}

Figure \ref{fig_map_eu} depicts heatmaps of the estimated tail (top) and scale parameter (bottom) on January~1, April~1, July~1 and October~1, from left to right. 
For most coordinate points, the tail parameter estimates are lowest on April~1 and highest on July~1, with 31.86\% of the estimates being equal to one on this day. For 632 out of the 656 coordinate points, the scale estimates are highest on July~1 and lowest on January~1. These patterns are also reflected in Table~\ref{sumstat1}, which reports summary statistics of the daily parameter estimates across meteorological seasons. Spatially, Figure~\ref{fig_map_eu} suggests that both the tail and the scale parameter values tend to be lower in the eastern part of the considered region \mbox{(20°~W\,-\,0°)}, which includes the British Isles and the northwest French coast. This indicates that the relative vorticity extremes behave differently, with stronger clustering near the coast and over land compared to further out in the North Atlantic.

\begin{table}[!htbp]
\centering
\caption{Summary statistics of the daily parameter estimates by meteorological season.}
\resizebox{\linewidth}{!}{
\begin{tabular}{cl|rrrrrrr}
  \hline
Parameter & Season & Min & 0.25 Quantile & Median  & 0.75 Quantile & Max & Mean & SD \\ 
  \hline
  $\beta$ & Winter & 0.611 & 0.766 & 0.802  & 0.841 & 0.995 & 0.805 & 0.053 \\ 
   & Spring & 0.482 & 0.694 & 0.733  & 0.775 & 1.000 & 0.738 & 0.069  \\ 
   & Summer & 0.476 &  0.841 & 0.922  & 1.000 & 1.000 & 0.901 & 0.099 \\ 
   & Fall & 0.595 & 0.800 &  0.843  & 0.893 & 1.000 & 0.847 & 0.068 \\ 
  \hline
  $\sigma$ & Winter & 128.4 & 266.1 & 288.4 & 312.3 & 508.9 & 289.8 & 34.9\\
  & Spring & 156.3 & 396.2 & 520.7 & 716.0 & 2273.8 & 589.9 & 259.1 \\
  & Summer & 292.3 & 737.3 & 911.1  & 1111.3 & 2743.6 & 942.9 & 279.3\\
  & Fall & 210.0 & 336.0 & 400.0  & 502.2 & 996.9 & 428.8 &  120.2\\
  \hline
\end{tabular}}
\label{sumstat1}
\end{table}

\begin{figure}[!htbp]
    \centering
    \includegraphics[width = 0.95\linewidth]{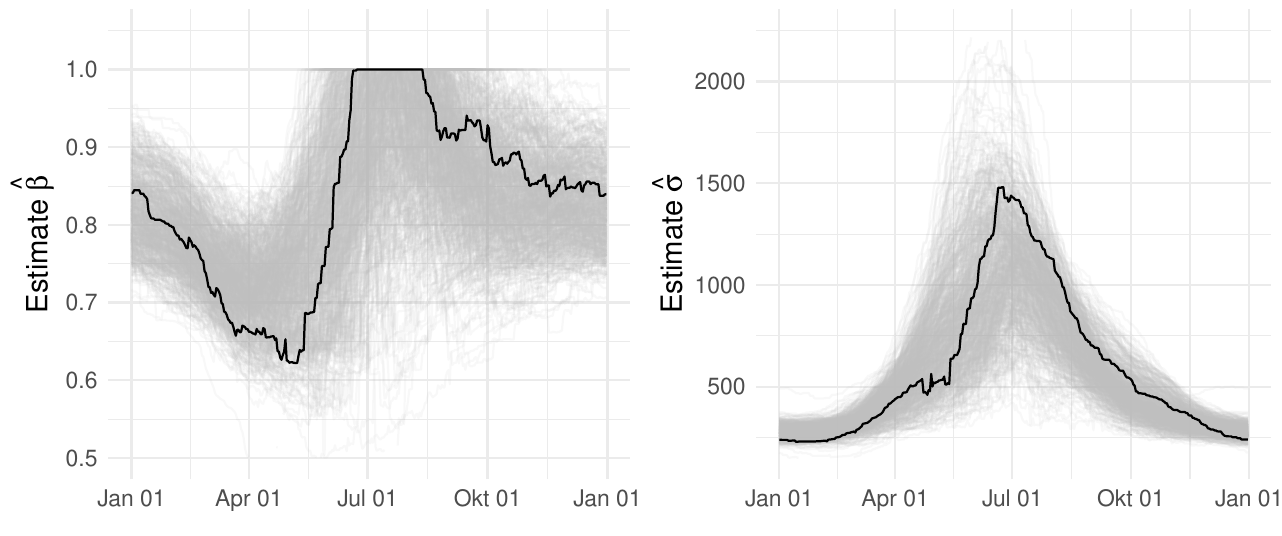}
    \caption{Annual course of the estimated tail parameter $\beta$ (left) and scale parameter $\sigma$ (right) of all coordinate points; estimates for \mbox{56°~N, 40°~W} are shown in black.}
    \label{fig_all_qb}
\end{figure}

Figure~\ref{fig_all_qb} shows the annual course of the parameter estimates for all 656 coordinate points, with the estimates for \mbox{56°~N, 40°~W} highlighted in black. This location was selected as an example because its parameter estimates are mostly centered within the range of estimated values across all coordinate points throughout the year. The figure illustrates that, at most coordinate points, the estimated tail parameter is lowest in April and May. This is consistent with the summary statistics in Table~\ref{sumstat1}, as the 0.25~quantile, median, mean and 0.75~quantile of the daily tail parameter estimates are lowest in spring. In contrast, the tail parameter estimates peak in July and August at most coordinate points, with 30.11\% being equal to one during these months.
This behavior is also reflected in Table~\ref{sumstat1}, as all summary statistics of the daily tail parameter estimates, except the minimum, are largest in summer.
For the scale parameter, it is evident from Figure~\ref{fig_all_qb} that the estimates are lower in boreal winter (DJF) and higher in boreal summer (JJA). Supporting this, each statistic in Table~\ref{sumstat1} has its lowest and highest values in winter and summer, respectively.

Figures \ref{fig_map_eu} and \ref{fig_all_qb} and Table~\ref{sumstat1} indicate that the deviation from an ordinary Poisson process is largest in boreal spring, suggesting that the clustering behavior of relative vorticity extremes is more pronounced during this season. In contrast, during summer, many tail parameter estimates are close to or equal to one, showing a behavior more similar to an ordinary Poisson process, particularly over the North Atlantic in the western part of the considered region \mbox{(40°~W\,-\,30°~W).}

\begin{figure}[!htbp]
    \centering
    \includegraphics[width = 0.95\linewidth]{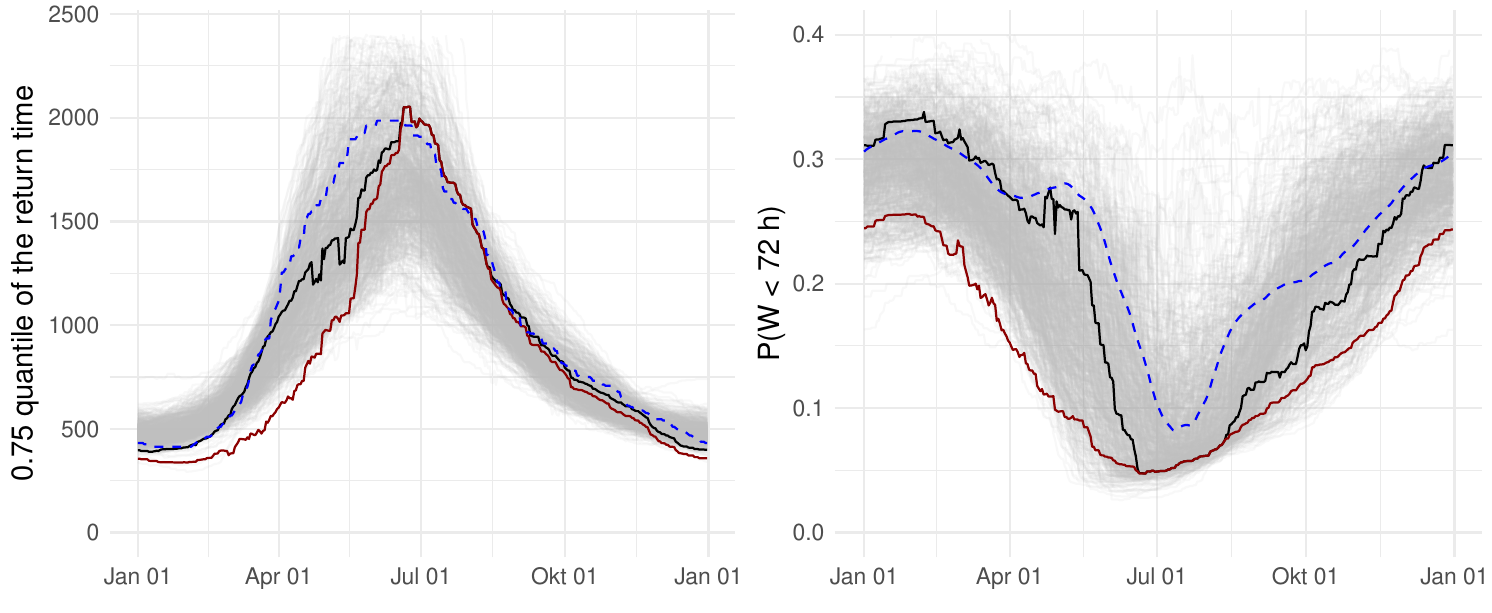}
    \caption{Annual course of the 0.75 quantile of the return time (left) and the estimated probability of at least one additional vorticity extreme occurring within the next 72~h (right) of all coordinate points; estimates for \mbox{56°~N, 40°~W} are shown in black (FPP) and red (ordinary PP); weighted empirical estimates are shown in blue with a dashed line.}
    \label{fig_med_p48}
\end{figure}

To translate the parameter estimates into application-oriented metrics, we examine the implied 0.75 quantile of the return time and the probability of at least one additional vorticity extreme occurring within the next 72 hours, shown in Figure~\ref{fig_med_p48}, again with the results for \mbox{56°~N, 40°~W} highlighted in black. Additionally, the results obtained using a seasonal ordinary Poisson process as model at \mbox{56°~N, 40°~W} are included in red. To evaluate which model better captures these metrics, the weighted empirical 0.75 quantile and the weighted relative frequency $h_{<\, 72}(j)$, given by 
\[h_{<\, 72}(j) = \sum_{i = 1}^n k_{c}(j, t_i) \cdot \mathds{1}\left(w_i < 72\right),\] are shown as blue dashed lines; notation as in Section~\ref{sec:methods_3}. 
The 0.75 quantile of the return time exhibits a similar pattern to the scale parameter estimates in Figure~\ref{fig_all_qb}. When comparing the red and the black lines, i.e., comparing the 0.75 quantile derived from an ordinary vs. fractional Poisson process, it becomes evident that, particularly in spring, the 0.75 quantile of the return times differ considerably. The weighted empirical 0.75 quantile is much more similar to the one derived from an FPP, with larger differences only occurring between mid-April and mid-June.  
For the estimated probability of at least one additional vorticity extreme occurring within the next 72 hours, the red and black lines substantially differ for most of the year.
Not only does the red line lie primarily below the black one, but it also lies below most of the other 655 lines, indicating that the ordinary Poisson process fails to capture the temporal clustering of vorticity extremes.
This interpretation is further supported by the weighted relative frequency $h_{<\, 72}(j)$, which is closer to the results obtained from an FPP throughout the entire year.
Nevertheless, from late spring to early fall, the FPP still appears to underestimate the probability of additional vorticity extremes occurring within 72 hours. 

Figure~\ref{fig_med_p48} illustrates that using the fractional Poisson process, and thereby including the additional tail parameter, affects not only the tails of the distribution but also properties such as the 0.75 quantile. For both properties considered here, at \mbox{56°~N, 40°~W}, kernel weighted daily empirical estimates are closer to the values derived from the FPP throughout the year. 
For this location specifically, the seasonal fractional Poisson process yields the lowest 0.75 quantile of the return time on January~10 and the highest probability of further vorticity extremes within the next 72 hours on February~7.

\begin{figure}[!htbp]
    \centering
    \includegraphics[width = 0.95\linewidth]{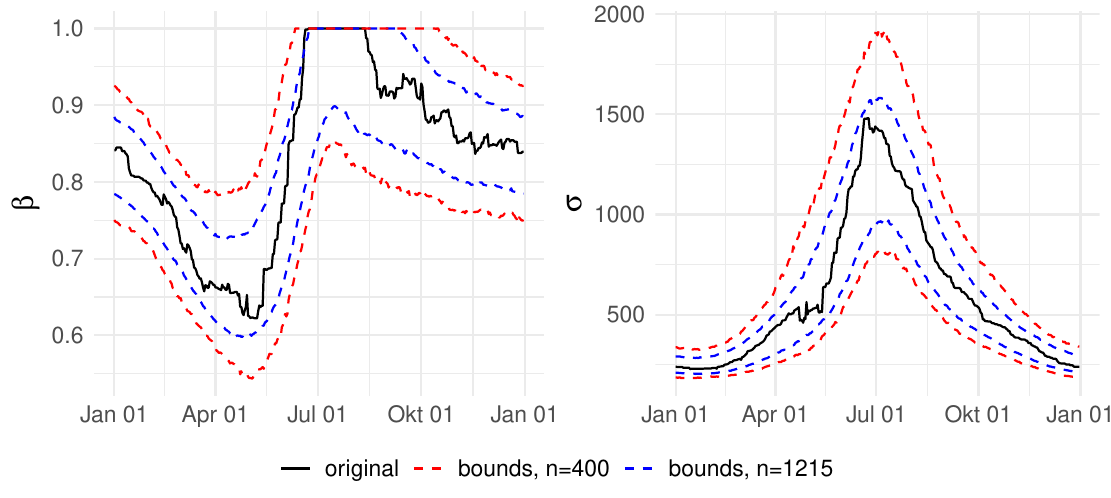}
    \caption{Underlying tail parameter $\beta$ (left) and scale parameter $\sigma$ (right) in black; 0.025 and 0.975 quantiles of the daily weighted quantile-based parameter estimates in red $(n = 400)$ and blue $(n = 1215)$.}
    \label{fig_ci}
\end{figure}

To assess the variability of the daily parameter estimates, we apply a parametric bootstrap procedure, i.e., we simulate return times from the distribution fitted at \mbox{56°~N, 40°~W} and calculate the daily weighted quantile-based parameter estimates for the simulated data. This procedure is repeated 1000 times for sample sizes $n = 1215$ (the number of return times per grid point present in the vorticity data) and $n = 400$ in order to also evaluate the effect of the sample size on the estimation quality. Then we compute the 0.025 and 0.975 quantiles of the 1000 estimates for each day and sample size, as shown in Figure~\ref{fig_ci}. 
For the tail parameter, the 0.025 quantile in summer lies above the 0.975 quantile in spring for both sample sizes, indicating a pronounced seasonal difference.
For the scale parameter, the 0.025 quantile in summer exceeds the 0.975 quantile in winter, as well as in substantial portions of spring and fall.
Furthermore, the parameter curves obtained for the real data remain relatively centered within the bounds for most of the year, except near points where the slope changes direction. This effect is most pronounced for the tail parameter between June~19 and 23, where the true parameter value even exceeds the upper bound for $n = 1215$.
For $n = 400$, the 0.025 and 0.975 quantiles are farther away from the true underlying parameter curves than for $n= 1215$, indicating lower estimation accuracy. However, this behavior is expected given that fewer than one third of the observations are available in this case.

	\section{Summary and Conclusion} \label{sec:conclusion}

The occurrences of vorticity extremes often display a clustering behavior with seasonal variations, which cannot be modeled adequately using an ordinary Poisson process.
Given the accumulated damage caused by their temporal clustering, it is crucial to account for this aspect in statistical modeling.
To capture this behavior, \citet{blender} proposed using the fractional Poisson process (FPP), a natural generalization of the ordinary Poisson process in which return times follow a heavy-tailed Mittag-Leffler distribution. However, due to the seasonality of vorticity extremes, they only considered data from boreal and austral winter.
Building on this approach, we extended the process by incorporating seasonality through periodically varying parameters in the Mittag-Leffler distribution, enabling year-round modeling. The time-dependent parameters are estimated separately for each calendar day using a distance-based weighting approach, with weights determined by an Epanechnikov kernel. 

In addition, we introduced a new quantile-based estimation approach for the Mittag-Leffler parameters, which minimizes the distance between the empirical and the theoretical Mittag-Leffler distribution at selected quantiles. This quantile-based estimation method, inspired by the Cramèr-von Mises minimization approach in \citet{christina}, outperforms the widely used method of log-moments in terms of MSE. While the maximum likelihood estimator achieves lower MSEs, it is considerably more computationally demanding. In our simulation study, the quantile-based estimator reached median empirical relative efficiencies of 0.76 for the tail parameter and 0.86 for the scale parameter compared to maximum likelihood, making it a strong and also robust alternative when computational resources are limited.

We applied the seasonal FPP to relative vorticity extremes in the North Atlantic-European Region using ERA5 data, where the extremes are determined with the peaks-over-threshold method. The vorticity extremes were modeled with the seasonal FPP for each grid point separately, yielding daily estimates of the tail and scale parameters. Across all coordinate points, the results indicate clustering behavior in winter, as well as most of spring and fall. In contrast, during summer, for a large fraction of the coordinate points, the parameter estimates pointed to a behavior as expected in an ordinary Poisson process. 
Furthermore, we compared the 0.75 quantile of the return time and the probability of additional vorticity extremes occurring within the next 72 hours obtained by an FPP and an ordinary Poisson process at \mbox{56° N\,-\,40° W}. In spring in particular, the ordinary Poisson process yielded lower 0.75 quantile estimates.  
Except for summer, the probability of at least one additional vorticity extreme occurring within the next 72 hours was estimated to be considerably lower with the ordinary Poisson process, indicating that it cannot adequately model the temporal clustering of vorticity extremes. A comparison with the weighted empirical 0.75 quantile and the weighted relative frequency suggests that the FPP was better able to capture these properties than the ordinary Poisson process.

    \section*{Acknowledgements}
    This research has been funded by the German Federal Ministry of Research, Technology and Space (BMFTR) within the subproject B3.3 (project number 01LP2323K) of the integrated project “Climate Change and Extreme Events – ClimXtreme Module B Statistics Phase II”. The authors gratefully acknowledge the computing time provided on the Linux HPC cluster at TU Dortmund University (LiDO3), partially funded in the course of the Large-Scale Equipment Initiative by the German Research Foundation (DFG) as project 271512359. The authors also express their gratitude to Christina Mathieu and Nils Weitzel, TU Dortmund University, and Joaquim Pinto, Karlsruhe Institute of Technology (KIT), for their valuable suggestions and insightful discussions. 
    During the preparation of this manuscript, we used the ChatGPT-4o model developed by OpenAI for minor language editing to improve readability. After using this tool, the authors carefully reviewed and edited the content as needed and take full responsibility for the content of the manuscript.

    \section*{Disclosure statement}
    The authors report that there are no competing interests to declare.
    
    \section*{Data Availability Statement}
    The R-code to reproduce the results is openly available at \url{https://github.com/smmlmend/quantile-based-estimation-Mittag-Leffler-distribution.git}. 
    The ERA5 reanalysis data are openly available through the Copernicus Climate Data Store at \url{https://cds.climate.copernicus.eu/datasets/derived-era5-pressure-levels-daily-statistics?tab=download}.

    \newpage
	\printbibliography[heading=bibintoc, title={References}]
    \newpage
    \appendix
    \numberwithin{equation}{section}
    \renewcommand{\theequation}{A-\arabic{equation}}

    \section{Appendix}
\subsection{Consistency and Asymptotic Normality of the Quantile-Based Estimator} \label{consistency}
     
In the following, we apply the implicit function theorem (see, e.g., \cite[Chapter~8]{implicit_fct_theorem}) together with the continuous mapping theorem (see, e.g., \cite[Chapter~1.7]{serfling}) to establish the consistency of the quantile-based estimator. 
Afterward, we use the delta method for implicitly defined random variables, introduced in \citet{implicit_delta}, to conclude asymptotic normality and construct asymptotic confidence intervals. Let 
     \begin{itemize}
         \item $F_{\theta}$ be the CDF of a continuous distribution with density $f_{\theta}$,
         \item$\alpha = (\alpha_1,\ldots, \alpha_{r})'$ with $0 < \alpha_1 < \ldots < \alpha_r < 1$ be a vector of probabilities,
         \item $\hat{q}_n = (\hat{q}_{n,1}, \ldots, \hat{q}_{n,r})'$ be the vector of the corresponding empirical $\alpha_1,\ldots,\alpha_r$ quantiles and
         \item $q_\theta = (q_{\theta,1}, \ldots, q_{\theta,r})'$ the theoretical $\alpha_1,\ldots,\alpha_r$ quantiles of $F_\theta$.
     \end{itemize}
     The quantile-based (QB) estimator is then given by 
     \[\hat{\theta}_n = \underset{ \theta}{\text{argmin}} \sum\limits_{i = 1}^r \bigl(\alpha_i - F_{\theta}(\hat{q}_{n,i}) \bigr) ^2.\]
Since $F_\theta$ is continuous and differentiable, the QB estimator $\hat{\theta}_n$ fulfills
    \[ 2 \sum\limits_{i = 1}^r \bigl(F_{\theta}(\hat{q}_{n,i})-
    \alpha_i \bigr) \cdot \frac{\partial}{\partial \theta}F_{\theta}(\hat{q}_{n,i})\Big|_{\theta = \hat{\theta}_n}  = 0.\]
In this section, we show that, for the Mittag-Leffler distribution and under mild conditions on $\alpha_1, \ldots, \alpha_r$, there exists a unique and continuously differentiable function $h$ satisfying $h(\hat{q}_n) = \hat{\theta}_n$ and $h(q_{\theta_0} )= \theta_0$, which then allows us to derive the consistency of $\hat{\theta}_n$ for the true parameter value $\theta_0 = (\beta_0, \sigma_0)' $.
For the Mittag-Leffler distribution, $\theta = (\beta , \sigma)' \in (0,1] \times \mathbb{R}^+$ with
 \[F_{\beta, \sigma}(x) = 1 - E_{\beta}\left(-\left(x/\sigma \right)^{\beta}\right) = 1-  \sum\limits_{k = 0}^\infty \frac{(-1)^k\left(\frac{x}{\sigma}\right)^{\beta k}}{\Gamma(1 + \beta k)} \quad \text{for } x > 0. \]
The density of the Mittag-Leffler distribution is given by
     \begin{align*}
        f_{\beta, \sigma}(x) &= \frac{\partial}{\partial x }F_{\beta, \sigma}(x) \\
        &= \frac{\partial}{\partial x } \left( 1- \sum\limits_{k = 0}^\infty
        \frac{(-1)^k\left( \frac{x}{\sigma}\right)^{\beta k}}{\Gamma(1+\beta k)}\right)\\
        &= -\sum\limits_{k = 1}^\infty \frac{(-1)^k \beta k}{\Gamma(1+\beta k)}\frac{1}{\sigma}\left(\frac{x}{\sigma}\right)^{\beta k-1} \\
        \end{align*}
        \begin{align*}
        \phantom{f_{\beta, \sigma}(x)} &=  \frac{1}{\sigma}\sum\limits_{k = 0}^\infty \frac{(-1)^k }{\Gamma(\beta k+\beta)}\left(\frac{x}{\sigma}\right)^{\beta k+\beta-1} \\  
        &= \frac{x^{\beta-1}}{\sigma^\beta} E_{\beta,\beta}\left(-\left(\frac{x}{\sigma}\right)^\beta\right).
    \end{align*}

 The implicit function theorem considers a function $h:\mathbb{R} ^{n}\to \mathbb {R} ^{m}$ whose graph $ ({\textbf {x}},h({\textbf {x}}))$ is precisely the set of all $(\textbf {x},\textbf {y})$ such that $ g({\textbf {x}},{\textbf {y}})={\textbf {0}}$ for a given function $g:\mathbb{R} ^{n+m}\to \mathbb {R} ^{m}$. Here, we use $\textbf {x} = q$ a vector of $r$ quantiles $q_1, \ldots, q_r$, $\textbf {y} = \theta = \left(\beta, \sigma\right)'$ and

  \begin{align*}
    g\left(q, \left(\beta, \sigma\right)\right)=  \left( \begin{array}{c} \frac{\partial}{\partial \beta}\sum\limits_{i = 1}^r \bigl(F_{\beta, \sigma}(q_i) -\alpha_i\bigr) ^2 \\  \frac{\partial}{\partial \sigma}\sum\limits_{i = 1}^r \bigl(F_{\beta, \sigma}(q_{i}) - \alpha_i \bigr) ^2 \end{array}\right)  = 
    \left( \begin{array}{c}2\sum\limits_{i = 1}^r(F_{\beta, \sigma}(q_{i})-\alpha_i)\cdot \frac{\partial}{\partial \beta}F_{\beta, \sigma}(q_{i}) \\  2\sum\limits_{i = 1}^r(F_{\beta, \sigma}(q_{i})-\alpha_i)\cdot \frac{\partial}{\partial \sigma} F_{\beta, \sigma}(q_{i}) \end{array} \right).
  \end{align*}

 The objective is to determine whether there exists a unique and continuously differentiable function $h$ such that $g(q, h(q)) = 0$. Note that $g(q_{\beta, \sigma}, (\beta, \sigma)) = 0$ holds for $q_{\beta, \sigma}$ the theoretical quantiles of $F_{\beta, \sigma}$, since $F_{\beta, \sigma}(q_{\beta, \sigma, i})-\alpha_i = 0 $ for $i = 1, \ldots, r$. By the implicit function theorem, the existence of $h$ is guaranteed if
  \[J_{g}\left(q, \left(\beta, \sigma \right)\right) =\begin{pmatrix}
  \frac{\partial}{\partial \beta} g_1\left(q, \left(\beta, \sigma\right)\right) &  \frac{\partial}{\partial \sigma} g_1\left(q, \left(\beta, \sigma\right)\right)\\
  \frac{\partial}{\partial \beta} g_2\left(q, \left(\beta, \sigma\right)\right)&  \frac{\partial}{\partial \sigma} g_2\left(q,\left(\beta, \sigma\right)\right)
 \end{pmatrix}\]
 is invertible at $(q_{\beta_0, \sigma_0}, \left(\beta_0, \sigma_0\right))$.
 This holds if
 {\footnotesize
\[\text{det}\bigl[ J_{g}\left(q_{\beta, \sigma}, (\beta, \sigma) \right) \bigr] = \frac{\partial}{\partial \beta} g_1(q_{\beta, \sigma},( \beta, \sigma)) \frac{\partial}{\partial \sigma} g_2(q_{\beta, \sigma}, (\beta, \sigma))  - \frac{\partial}{\partial \sigma} g_1(q_{\beta, \sigma},( \beta, \sigma))\frac{\partial}{\partial \beta} g_2(q_{\beta, \sigma}, (\beta, \sigma)) \neq 0\]}at $(q_{\beta_0, \sigma_0}, \left(\beta_0, \sigma_0\right))$. After lengthy calculations, we obtain
 \begin{align}\label{det}
     \text{det}\bigl[ J_{g}\left(q_{\beta_0, \sigma_0}, (\beta_0, \sigma_0) \right) \bigr]
     = 2  \sum \limits_{i = 1}^r \sum \limits_{j = 1}^r & \left( \frac{\partial}{\partial \sigma} F_{\beta_0,\sigma_0}(q_{\beta_0, \sigma_0,i})  \frac{\partial}{\partial \beta } F_{\beta_0,\sigma_0}(q_{\beta_0, \sigma_0,j}) \right.  \\
     & \left. \quad \, \; \, - \frac{\partial}{\partial \sigma } F_{\beta_0,\sigma_0}(q_{\beta_0, \sigma_0,j})  \frac{\partial}{\partial \beta } F_{\beta_0,\sigma_0}(q_{\beta_0, \sigma_0,i})  \right)^2. \nonumber
    \end{align}
Hence, $ \text{det}\bigl[ J_{g}\left(q_{\beta_0, \sigma_0}, (\beta_0, \sigma_0) \right) \bigr]\geq 0$ for all $\beta_0 \in (0,1]$, $\sigma_0 > 0 $ and choices of $\alpha_1, \ldots, \alpha_r$. Moreover, whenever
\begin{align}\label{A1}
\frac{\partial}{\partial \sigma } F_{\beta_0,\sigma_0}(q_{\beta_0, \sigma_0, i})  \frac{\partial}{\partial \beta } F_{\beta_0,\sigma_0}(q_{\beta_0, \sigma_0, j}) \neq  \frac{\partial}{\partial \sigma} F_{\beta_0,\sigma_0}(q_{\beta_0, \sigma_0, j})  \frac{\partial}{\partial \beta} F_{\beta_0,\sigma_0}(q_{\beta_0, \sigma_0, i})\end{align} 
holds for at least one pair of $i \neq j$, we have $ \text{det}\bigl[ J_{g}\left(q_{\beta, \sigma}, (\beta, \sigma) \right) \bigr]> 0$. 
Calculating the partial derivative of $F_{\beta, \sigma}$ with respect to $\sigma$ yields
\[ \frac{\partial}{\partial \sigma } F_{\beta,\sigma}(x) =- \frac{x}{\sigma} f_{\beta,\sigma}(x),\] which is negative for all $x > 0$.
The partial derivative of $F_{\beta, \sigma}$ with respect to $\beta$ is
\[\frac{\partial}{\partial \beta } F_{\beta,\sigma}(x)= -\sum\limits_{k=1}^\infty (-1)^k k \left( \frac{x}{\sigma}\right)^{\beta k}\frac{ \ln\left( \frac{x}{\sigma}\right) -  \psi(1+\beta k)}{\Gamma(1+\beta k)},\] 
where $\psi$ denotes the digamma function. The derivative $\frac{\partial}{\partial \beta } F_{\beta,\sigma}(x)$ is proven to be negative for sufficiently small $x$, whereas asymptotic considerations and numerical evidence strongly suggest that it becomes positive for sufficiently large $x$ (for details, see Subsection~\ref{sign_derivative}). Therefore, selecting at least one sufficiently small and one sufficiently large quantile ensures the inequality in (\ref{A1}) due to the difference in sign. Consequently, $ J_{g}\left(q_{\beta, \sigma}, (\beta, \sigma) \right) $ is invertible, and the conditions of the implicit function theorem are satisfied.

As $E_{\beta, \beta}(z)$ has no negative zeros for $\beta \in (0,1]$ \citep[Chapter~4.6]{Gorenflo}, the density $f_{\beta, \sigma}$ is strictly positive and consequently the CDF is strictly increasing. Hence, the empirical quantiles are strongly consistent estimators of the theoretical quantiles \citep[Chapter~2.3]{serfling}, i.e.,
\[  \hat{q}_{n}  \overset{\text{a.s.}\;}{\longrightarrow}  q_{\beta, \sigma}  . \]
Since the implicit function theorem guarantees the existence of a unique and continuously differentiable function $h$ for appropriately selected quantiles, the continuous mapping theorem implies that
\[ h(\hat{q}_{n})  \overset{\text{a.s.}\;}{\longrightarrow}  h(q_{\beta, \sigma}) = (\beta, \sigma)'. \]
Therefore, the quantile-based estimator is consistent whenever the quantiles are selected such that (\ref{A1}) holds for at least one quantile pair. 

In the following, we establish the asymptotic normality of the quantile-based estimator under conditions on the quantile selection similar to those required for consistency. The empirical quantiles are asymptotically normal (see, e.g., \cite[Chapter~2.3]{serfling}) with
\[  \sqrt{n}(\hat{q}_{n} -q_{\beta, \sigma})  \sim \mathcal{AN}\left( 0 , \Sigma\right) \quad \text{and} \quad  \Sigma_{ij} =  \frac{\alpha_i (1- \alpha_j)}{f(q_{\beta, \sigma,i})f(q_{\beta, \sigma,j})} \quad \text{for }i\leq j\] and $\Sigma_{ij} = \Sigma_{ji}$ for $i > j$. Then, under the conditions required for consistency above and one additional condition, the delta method for implicitly defined random variables states
\[ \sqrt{n} ((\hat{\beta}_n, \hat{\sigma}_n)' - (\beta, \sigma)') \sim \mathcal{AN}(0, J^{-1}H\Sigma H'(J^{-1})'), \]
where
\[J = J_{g}\left(q_{\beta, \sigma}, (\beta, \sigma) \right) \quad \text{and} \quad H = \left(\frac{\partial}{\partial q_i} g_\ell \left(q_{\beta, \sigma}, \left(\beta, \sigma\right)\right) \right)_{\ell = 1,2,\,i = 1, \ldots, r}\]
are evaluated at $(q_{\beta_0, \sigma_0}, \left(\beta_0, \sigma_0\right))$ with $J_g$ and $g()$ as previously defined. 
The additional condition needed to apply the delta method for implicitly defined random variables is that each row of the matrix $J^{-1} H$ contains at least one nonzero element. Lengthy calculations yield that this is fulfilled if each of the following inequalities holds for at least one quantile $q_{\beta_0, \sigma_0, j}$.
\begin{align}
    \frac{\partial}{\partial \sigma}F(q_{\beta_0, \sigma_0, j}) \sum_{i = 1}^r \left(\frac{\partial}{\partial \beta}F(q_{\beta_0, \sigma_0, i})\right)^2 \neq \frac{\partial}{\partial \beta}F(q_{\beta_0, \sigma_0, j}) \sum_{i = 1}^r \frac{\partial}{\partial \beta}F(q_{\beta_0, \sigma_0, i})\frac{\partial}{\partial \sigma}F(q_{\beta_0, \sigma_0, i}), \label{ineq1}\\
    \frac{\partial}{\partial \beta}F(q_{\beta_0, \sigma_0, j}) \sum_{i = 1}^r \left(\frac{\partial}{\partial \sigma}F(q_{\beta_0, \sigma_0, i})\right)^2 \neq \frac{\partial}{\partial \sigma}F(q_{\beta_0, \sigma_0, j}) \sum_{i = 1}^r \frac{\partial}{\partial \beta}F(q_{\beta_0, \sigma_0, i})\frac{\partial}{\partial \sigma}F(q_{\beta_0, \sigma_0, i}). \label{ineq2}
\end{align}
Given that \[\sum_{i = 1}^r \left(\frac{\partial}{\partial \beta}F(q_{\beta_0, \sigma_0, i})\right)^2 > 0, \quad \sum_{i = 1}^r \left(\frac{\partial}{\partial \sigma}F(q_{\beta_0, \sigma_0, i})\right)^2 > 0 \quad \text{and} \quad \frac{\partial}{\partial \sigma}F(q_{\beta_0, \sigma_0, j}) < 0,\]
the inequalities can be ensured by selecting $q_{\beta_0, \sigma_0, j}$ such that the two sides of the inequalities have different signs. For this purpose, inequalities (\ref{ineq1}) and (\ref{ineq2}) can be divided into three cases 
\[\sum_{i = 1}^r \frac{\partial}{\partial \beta}F(q_{\beta_0, \sigma_0, i})\frac{\partial}{\partial \sigma}F(q_{\beta_0, \sigma_0, i}) \left\{
\begin{array}{ll}
 \; > 0, \quad (\text{I})\\
 \; < 0, \quad (\text{II})\\
  \;= 0.\quad (\text{III})
\end{array}\right.\]
In case (I), both inequalities are satisfied if at least one $q_{\beta_0, \sigma_0, j}$ is chosen sufficiently large such that $\frac{\partial}{\partial \beta}F(q_{\beta_0, \sigma_0, j})> 0 $. In case (II), the inequalities hold if at least one $q_{\beta_0, \sigma_0, j}$ is chosen sufficiently small such that $\frac{\partial}{\partial \beta}F(q_{\beta_0, \sigma_0, j})< 0$.
In case (III), the right-hand side of both inequalities reduces to zero. Hence, the inequalities are fulfilled provided that at least one $q_{\beta_0, \sigma_0, j}$ is chosen sufficiently large or small so that $\frac{\partial}{\partial \beta}F(q_{\beta_0, \sigma_0, j})\neq 0$.
Consequently, selecting at least one sufficiently small and one sufficiently large quantile ensures that inequalities (\ref{ineq1}) and (\ref{ineq2}) are satisfied in all three cases, which allows applying the delta method for implicitly defined random variables.
Therefore, under an appropriate selection of quantiles, the quantile-based estimator is consistent and asymptotically normal.

Under an appropriate selection of quantiles ensuring asymptotic normality of the quantile-based estimators, approximate $(1-\tilde{\alpha})$ confidence intervals for $\beta$ and $\sigma$ are given by
\begin{equation}
\left[\max\left\{0, \hat{\beta}_n - z_{1- \frac{\tilde{\alpha}}{2}}\sqrt{\frac{\widehat{V}_{11}}{n}}\right\} \, , \, \min\left\{1, \hat{\beta}_n + z_{1- \frac{\tilde{\alpha}}{2}}\sqrt{\frac{\widehat{V}_{11}}{n}}\right\}\right] \tag{CI-1}
\label{CI1}\end{equation}
and
\begin{equation}\left[ \max \left\{0,  \hat{\sigma}_n - z_{1- \frac{\tilde{\alpha}}{2}}\sqrt{\frac{\widehat{V}_{22}}{n}} \right\} \, , \,  \hat{\sigma}_n + z_{1- \frac{\tilde{\alpha}}{2}}\sqrt{\frac{\widehat{V}_{22}}{n}} \right], \tag{CI-2}
\label{CI2}\end{equation}
where $z_{1- \frac{\tilde{\alpha}}{2}}$ denotes the $(1- \frac{\tilde{\alpha}}{2})$ quantile of the standard normal distribution. $\widehat{V}_{11}$ and $\widehat{V}_{22}$ are estimators of the corresponding diagonal elements of the matrix $J^{-1}H\Sigma H'(J^{-1})'$, obtained by replacing the true underlying parameters $\beta$ and $\sigma$ with the QB estimators $\hat{\beta}_n$ and $\hat{\sigma}_n$.

To assess whether the confidence intervals achieve the nominal coverage probability, we conduct a simulation study. We consider tail parameter values $\beta = 0.7, 0.95 $, scale parameter values $\sigma = 5, 250 $, sample sizes $n = 50, 250, 1000, 5000$ and the quantile combinations
\[\text{Comb 1: }(0.1, 0.3, 0.5, 0.8, 0.925) \quad \text{and} \quad  \text{Comb 2: } (0.05, 0.2, 0.475, 0.8, 0.925).\]
The quantile combinations were selected based on their favorable performance with respect to criteria~(\ref{comp}) and (\ref{extra}) and the Mittag-Leffler parameters were selected based on the results of the application to relative vorticity extremes in Section~\ref{sec:example}.
For each combination of sample size, tail parameter and scale parameter, 10,000 datasets are generated from the corresponding Mittag-Leffler distribution. The empirical coverage rate is then computed for each setting separately as the proportion of datasets for which the true parameters fall within the 0.95~confidence intervals. The empirical coverage rates and the mean confidence intervals lengths are reported in Table~\ref{coverage}. 

\begin{table}[!htbp]
\centering
\caption{Empirical coverage rates and mean interval lengths of the confidence intervals in (\ref{CI1}) and (\ref{CI2}) in the simulation study with a nominal coverage probability of 0.95.}
\begin{tabular}{rrr|cc|cc|cc|cc}
  \hline
 &&& \multicolumn{4}{c}{Comb 1} & \multicolumn{4}{c}{Comb 2} \\ 
  && &  \multicolumn{2}{c}{$\beta$}  & \multicolumn{2}{c}{$\sigma$} & \multicolumn{2}{c}{$\beta$}  & \multicolumn{2}{c}{$\sigma$ }\\
  $n$ & $\beta$ & $\sigma$ & rate & length & rate & length & rate & length & rate & length\\
  \hline
   50 & 0.7 & 5 & 0.916 & 0.271 
   & 0.923 &  \phantom{11}5.90
   & 0.919 & 0.271
   & 0.927 & \phantom{11}5.93\\
   & & 250 &  0.917 &  0.272&
   0.929 &  295.88 &
   0.921 &  0.271 & 0.9332
   & 295.32\\
    & 0.95 & 5 & 0.991 & 0.182
    & 0.941 & \phantom{11}3.59
    & 0.993 & 0.179
    & 0.941 & \phantom{11}3.59\\ 
    & & 250 &  0.991 & 0.182
    & 0.941 & 179.55
    & 0.993 & 0.179
    & 0.941 & 179.56 \\
  \hline
   250 & 0.7 & 5 & 0.942 & 0.123
    & 0.944 & \phantom{11}2.63
    & 0.939 & 0.122
    & 0.944  & \phantom{11}2.65\\
    & & 250 & 0.942 & 0.123
    & 0.943 & 131.79
    & 0.939 & 0.122
    & 0.941 & 132.60\\
    & 0.95 &  5 & 0.983 & 0.102
    & 0.947 & \phantom{11}1.61
    & 0.986 & 0.101
    &  0.950 & \phantom{11}1.61\\
    & & 250 & 0.983 & 0.102
    & 0.947 & \phantom{1}80.45
    & 0.985 & 0.102
    & 0.949 & \phantom{1}80.69\\
  \hline
   1000 & 0.7 & 5 & 0.947 &  0.061
    &  0.945 & \phantom{11}1.32
    & 0.946 &    0.061
    & 0.944& \phantom{11}1.33 \\
    & & 250 &  0.947 & 0.061
    &  0.940 & \phantom{1}66.02
    &  0.946 & 0.061
    & 0.938 & \phantom{1}66.45\\
    & 0.95 & 5 & 0.949 & 0.061
    & 0.945 & \phantom{11}0.81
    & 0.948 & 0.061
    & 0.943 & \phantom{11}0.81\\
    & & 250 & 0.949 & 0.061
    & 0.942 & \phantom{1}40.47
    & 0.947 &  0.061
    &  0.942& \phantom{1}40.62 \\
  \hline
    5000 & 0.7 & 5 & 0.951 & 0.027
    & 0.950 & \phantom{11}0.59
    & 0.952 & 0.027
    & 0.949  & \phantom{11}0.59\\
    & &  250 & 0.951 & 0.027
    & 0.944 & \phantom{1}29.52
    & 0.954 & 0.027
    &  0.945 & \phantom{1}29.72\\
    & 0.95 & 5 & 0.951 & 0.028
    &  0.947 & \phantom{11}0.36
    & 0.950 & 0.028
    & 0.949 & \phantom{11}0.36 \\
    & & 250 & 0.951 & 0.028
    & 0.945 & \phantom{1}18.12
    & 0.949 & 0.028
    & 0.944 & \phantom{1}18.19 \\
    \hline
\end{tabular}
\label{coverage}
\end{table}

For $n = 50$ and $250$ the empirical coverage rates of the confidence interval for $\beta$ in (\ref{CI1}) deviate somewhat from the nominal coverage probability. However, this effect vanishes for larger sample sizes ($n = 1000, 5000$), where the intervals provide quite reliable coverage. The scale parameter lies outside the confidence interval in (\ref{CI2}) slightly too often, especially for $n = 50$, however, this is less pronounced for $n = 5000$, with empirical coverage rates between $0.944$ and $0.950$. For all sample sizes except $n = 50$, the empirical coverage rates of the interval in (\ref{CI2}) are closer to the nominal level for $\sigma = 5$ than for $\sigma = 250$. This effect is due to small numerical instabilities that arise when computing $\widehat{V}_{22}$ for large values of $\sigma$, and can be circumvented by rescaling the data. The confidence interval lengths decrease with increasing sample size, as expected approximately at a rate proportional to $\frac{1}{\sqrt{n}}$. Also, the lengths depend on the magnitude of the underlying parameter, with $\sigma = 250$ yielding much longer intervals than $\sigma = 5$. Furthermore, truncation at 0 or 1 in (\ref{CI1}) seems to affect the interval lengths, as $\beta = 0.95$ consistently produces shorter intervals than $\beta = 0.7$. Overall, the two quantile combinations Comb~1 and Comb~2 yield comparable results.

\subsection[Sign of $\frac{\partial}{\partial \beta } F_{\beta,\sigma}(x)$]{Sign of $\boldsymbol{\frac{\partial}{\partial \beta } F_{\beta,\sigma}(x)}$}
\label{sign_derivative}

Recently, \citet{monotonicity} showed that for any $\beta_0 \in(0, 1)$ and 
\begin{align} \label{interval}
 x \in \left(0, \min\left(\exp\left(-\frac{7}{2} \right),  2^{-\frac{1}{\beta_0}}\right) \right],\end{align} the Mittag-Leffler
function $ E_{\beta}(-x^{\beta})$ is monotonically increasing in $\beta \in [\beta_0, 1]$. Note that $\beta_0$ here is not the same as in Section~\ref{consistency}.
This result implies \[\frac{\partial}{\partial \beta } F_{\beta,\sigma}(x) = \frac{\partial}{\partial \beta }\left[1- E_{\beta}\left(-\left(\frac{x}{\sigma}\right)^{\beta}\right) \right]< 0 \text{  for  } \frac{x}{\sigma} \in \left(0, \min\left( \exp\left(-\frac{7}{2} \right), 2^{-\frac{1}{\beta_0}} \right) \right]\]
and $\beta \in [\beta_0, 1]$. Examining the upper limit of the interval in (\ref{interval}) yields
\[F_{\beta_0,1}\left(\min\left( \exp\left(-\frac{7}{2} \right), 2^{-\frac{1}{\beta_0}} \right)\right)\geq 0.0297 \quad \text{for all} \quad \beta_0 \in (0, 1). \] From this, we conclude that \[\frac{\partial}{\partial \beta } F_{\beta,\sigma}(q_{\alpha, \beta, \sigma}) = \frac{\partial}{\partial \beta } F_{\beta,1}(q_{\alpha, \beta, 1}) < 0\] holds for all $\beta \in (0, 1]$ and $\sigma > 0$ if $\alpha < 0.0297 $, which is rather restrictive. However, by extending the upper limit of interval~(\ref{interval}), we show that $\frac{\partial}{\partial \beta } F_{\beta,\sigma}(q_{\alpha, \beta, \sigma})  < 0$ for $\alpha < 0.1797 $.

\begin{figure}[!htbp]
    \centering
    \includegraphics[width=0.7\linewidth]{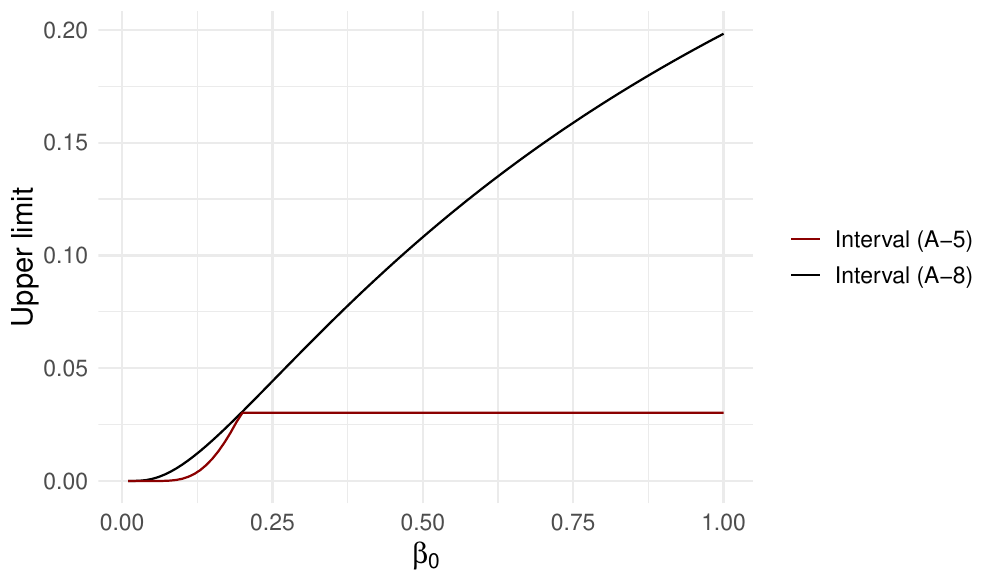}
    \caption{Upper limits of interval~(\ref{interval}) and interval~(\ref{intervalz_sup}).}
    \label{fig:upper_limit}
\end{figure}

In their proof, \citet{monotonicity} consider
\[\frac{\partial}{\partial \beta } E_{\beta}(-x^\beta)= \sum\limits_{k=1}^\infty (-1)^k k x^{\beta k}\frac{ \ln\left( x\right) -  \psi(1+\beta k)}{\Gamma(1+\beta k)} = \sum\limits_{k=1}^\infty (-1)^k y_k\]
and examine
\[ \frac{y_{k+1}}{y_k} = x^\beta \frac{\Gamma (\beta k )}{\Gamma (\beta k + \beta )} \left( 1 + \frac{ \psi(1+\beta k) -  \psi(1+ \beta + \beta k)}{\ln(x) -  \psi(1+ \beta k)}\right).\]
They show that
\begin{align} \label{bound1}
    \left|\frac{ \psi(1+\beta k) -  \psi(1+ \beta + \beta k)}{\ln(x) -  \psi(1+ \beta k) } \right| \leq \frac{\frac{5}{2}}{| \ln(x)| -1 }
\end{align} 
holds, which is less than $1$ for $ x < \exp\left(-\frac{7}{2}\right)$. Then 
\[\frac{y_{k+1}}{y_k} \leq  x^\beta \left( 1 + \frac{ \psi(1+\beta k) -  \psi(1+ \beta + \beta k)}{\ln(x) -  \psi(1+ \beta k)}\right) < 2x^\beta, \]
which is less than $1$ for $x < 2^{-\frac{1}{\beta}}$. Since $y_k < 0 $ for $x < \exp(\psi(1))$ and $\exp\left(-\frac{7}{2} \right) < \exp(\psi(1))$, selecting $x$ within interval (\ref{interval}) yields $y_{k+1} > y_k$ and 
\[ \frac{\partial}{\partial \beta } E_{\beta}(-x^\beta)=  \sum\limits_{k=1}^\infty (-1)^k y_k = -(y_1 -y_2) - (y_3- y_4) - \ldots > 0. \]
See \citet{monotonicity} for the preceding considerations. In the following, we increase interval~(\ref{interval}).
Let $x < \exp(-1) < \exp(\psi(1))$, such that $y_k < 0$ and 
\[\frac{ \psi(1+\beta k) -  \psi(1+ \beta + \beta k)}{\ln(x) -  \psi(1+ \beta k)} > 0\]
hold, as both the denominator and the numerator are negative. We extend interval (\ref{interval}) by replacing $1$, used as the upper bound in (\ref{bound1}) in the previous work, with $z > 0$ instead. This yields
\[ \frac{ \psi(1+\beta k) -  \psi(1+ \beta + \beta k)}{\ln(x) -  \psi(1+ \beta k)} \leq \frac{\frac{5}{2}}{| \ln(x)| -1 } < z \text{ for } x < \exp\left(-\frac{5}{2}\cdot \frac{1}{z} -1 \right). \]
This implies
\[\frac{y_{k+1}}{y_k}\leq  x^\beta \left( 1 + \frac{ \psi(1+\beta k) -  \psi(1+ \beta + \beta k)}{\ln(x) -  \psi(1+ \beta k)}\right)  < x^\beta(z+1), \]
which is less than 1 for $ x < (z+1)^{-\frac{1}{\beta}}$. Hence, selecting 
\begin{align} \label{intervalz}
    x \in \left(0, \min\left( \exp\left(-\frac{5}{2}\frac{1}{z}-1 \right), (z+1)^{-\frac{1}{\beta_0}}  \right)\right),\end{align} with $z >0 $, ensures that $\frac{\partial}{\partial \beta } E_{\beta}(-x^\beta) > 0$ for $\beta \in [\beta_0, 1]$. Note that for $z = 1$, the interval in (\ref{intervalz}) coincides with the interval found by \citet{monotonicity}.
   \newline Since $\exp\left(-\frac{5}{2}\frac{1}{z}-1 \right) \in \left(0, \exp(-1)\right]$ is monotonically increasing and $ (z+1)^{-\frac{1}{\beta}}\in (0, 1]$ is monotonically decreasing in $z > 0$, the upper limit in (\ref{intervalz}) is maximized for the value $z_0$ satisfying \[\exp\left(-\frac{5}{2}\frac{1}{z_0}-1 \right) =  (z_0+1)^{-\frac{1}{\beta_0}}.\]
Therefore, for each $\beta_0\in (0,1)$, the interval for $x$ in which $ E_{\beta}(-x^{\beta})$ is monotonically increasing in $\beta \in [\beta_0, 1]$ is given by
\begin{align} \label{intervalz_sup} x \in \left(0, \underset{z > 0 }{\sup}\left\{ \min\left( \exp\left(-\frac{5}{2}\frac{1}{z}-1 \right), (z+1)^{-\frac{1}{\beta_0}}  \right) \right\} \right).\end{align}
Consequently, $F_{\beta, \sigma}(x)$ is monotonically decreasing whenever $\frac{x}{\sigma}$ lies within this interval. The upper limits of interval~(\ref{intervalz_sup}) and interval~(\ref{interval}) are displayed in Figure~\ref{fig:upper_limit}. From the upper limit of interval~(\ref{intervalz_sup}), we can derive that $\frac{\partial}{\partial \beta } F_{\beta,\sigma}(q_{\alpha, \beta, \sigma})  < 0$ for $\alpha < 0.1797 $, since\[F_{\beta_0,1}\left(\underset{z > 0 }{\sup}\left\{ \min\left( \exp\left(-\frac{5}{2}\frac{1}{z}-1 \right), (z+1)^{-\frac{1}{\beta_0}}  \right) \right\} \right)\geq 0.1797 \quad \text{for all} \quad \beta_0 \in (0, 1). \]

Additionally, we conjecture that $\frac{\partial}{\partial \beta } F_{\beta,\sigma}(x) > 0$ holds for sufficiently large $x$. 
To the best of our knowledge, this has not yet been shown analytically. However, we examined the behavior of $\frac{\partial}{\partial \beta } F_{\beta,\sigma}(x)$ numerically on a grid, varying both $\alpha$ and $\beta$ in increments of 0.0001 starting at 0.0001 and ending at 0.9999 and 1, respectively. We found that
$\frac{\partial}{\partial \beta } F_{\beta,\sigma}(q_p) > 0$ for all $\beta \in (0, 1]$ whenever $\alpha > 0.5935$.
This also makes sense considering the asymptotic behavior of $E_\beta (-x^\beta)$, since 
\[ E_\beta(-x^\beta) \sim \frac{x^{-\beta}}{\Gamma(1-\beta)} \quad \text{for}\quad x \rightarrow \infty,\]which is monotonically decreasing in $\beta$ \citep{Mainardi_2014}, implying an increasing behavior for $F_{\beta,1}(x) = 1- E_\beta(-x^\beta)$.

\newpage

\subsection{Simulation Study}

\begin{table}[ht]
\centering
\caption{The 25 quantile combinations with the best performance in terms of criterion (\ref{comp}) and their performance according to criterion (\ref{extra}).}
\begin{tabular}{cccccccc}
  \hline
$\alpha_1$ & $\alpha_2$ & $\alpha_3$ & $\alpha_4$ & $\alpha_5$ &  criterion (\ref{comp})  &  criterion (\ref{extra}) & rank with (\ref{extra}) \\ 
  \hline
  0.100 & 0.300 & 0.500 & 0.800 & 0.925 & 1.503 & 0.581 & \phantom{0}8\\ 
  0.100 & 0.300 & 0.525 & 0.800 & 0.925 & 1.503 & 0.581 & \phantom{0}7 \\ 
  0.050 & 0.200 & 0.475 & 0.800 & 0.925 & 1.503 & 0.584 & \phantom{0}1 \\ 
  0.050 & 0.200 & 0.500 & 0.800 & 0.925 & 1.502 & 0.583 & \phantom{0}3 \\ 
  0.100 & 0.300 & 0.550 & 0.800 & 0.925 & 1.502 & 0.579 & 13 \\ 
  0.050 & 0.200 & 0.450 & 0.800 & 0.925 & 1.501 & 0.583 & \phantom{0}2 \\ 
  0.075 & 0.225 & 0.500 & 0.800 & 0.925 & 1.501 & 0.581 & \phantom{0}9 \\ 
  0.075 & 0.225 & 0.475 & 0.800 & 0.925 & 1.501 & 0.581 & \phantom{0}6 \\ 
  0.050 & 0.225 & 0.500 & 0.800 & 0.925 & 1.500 & 0.580 & 11 \\ 
  0.075 & 0.200 & 0.475 & 0.800 & 0.925 & 1.500 & 0.582 & \phantom{0}5 \\ 
  0.100 & 0.275 & 0.500 & 0.800 & 0.925 & 1.500 & 0.579 & 17 \\ 
  0.125 & 0.300 & 0.525 & 0.800 & 0.925 & 1.500 & 0.577 & 26 \\ 
  0.075 & 0.200 & 0.450 & 0.800 & 0.925 & 1.500 & 0.582 & \phantom{0}4 \\ 
  0.125 & 0.300 & 0.500 & 0.800 & 0.925 & 1.499 & 0.577 & 28 \\ 
  0.050 & 0.225 & 0.475 & 0.800 & 0.925 & 1.499 & 0.580 & 10 \\ 
  0.075 & 0.300 & 0.550 & 0.800 & 0.925 & 1.499 & 0.577 & 25 \\ 
  0.075 & 0.200 & 0.500 & 0.800 & 0.925 & 1.499 & 0.580 & 12 \\ 
  0.125 & 0.300 & 0.550 & 0.800 & 0.925 & 1.499 & 0.575 & 41 \\ 
  0.100 & 0.275 & 0.525 & 0.800 & 0.925 & 1.498 & 0.577 & 19 \\ 
  0.075 & 0.250 & 0.500 & 0.800 & 0.925 & 1.498 & 0.577 & 21 \\ 
  0.100 & 0.250 & 0.500 & 0.800 & 0.925 & 1.498 & 0.577 & 20 \\ 
  0.075 & 0.300 & 0.525 & 0.800 & 0.925 & 1.498 & 0.576 & 35 \\ 
  0.100 & 0.225 & 0.500 & 0.800 & 0.925 & 1.497 & 0.578 & 18 \\ 
  0.075 & 0.275 & 0.500 & 0.800 & 0.925 & 1.497 & 0.576 & 33 \\ 
  0.050 & 0.200 & 0.525 & 0.800 & 0.925 & 1.497 & 0.579 & 15 \\
   \hline
\end{tabular}
\label{25comb}
\end{table}

\begin{figure}[ht]
    \centering
    \includegraphics[width=0.92\linewidth]{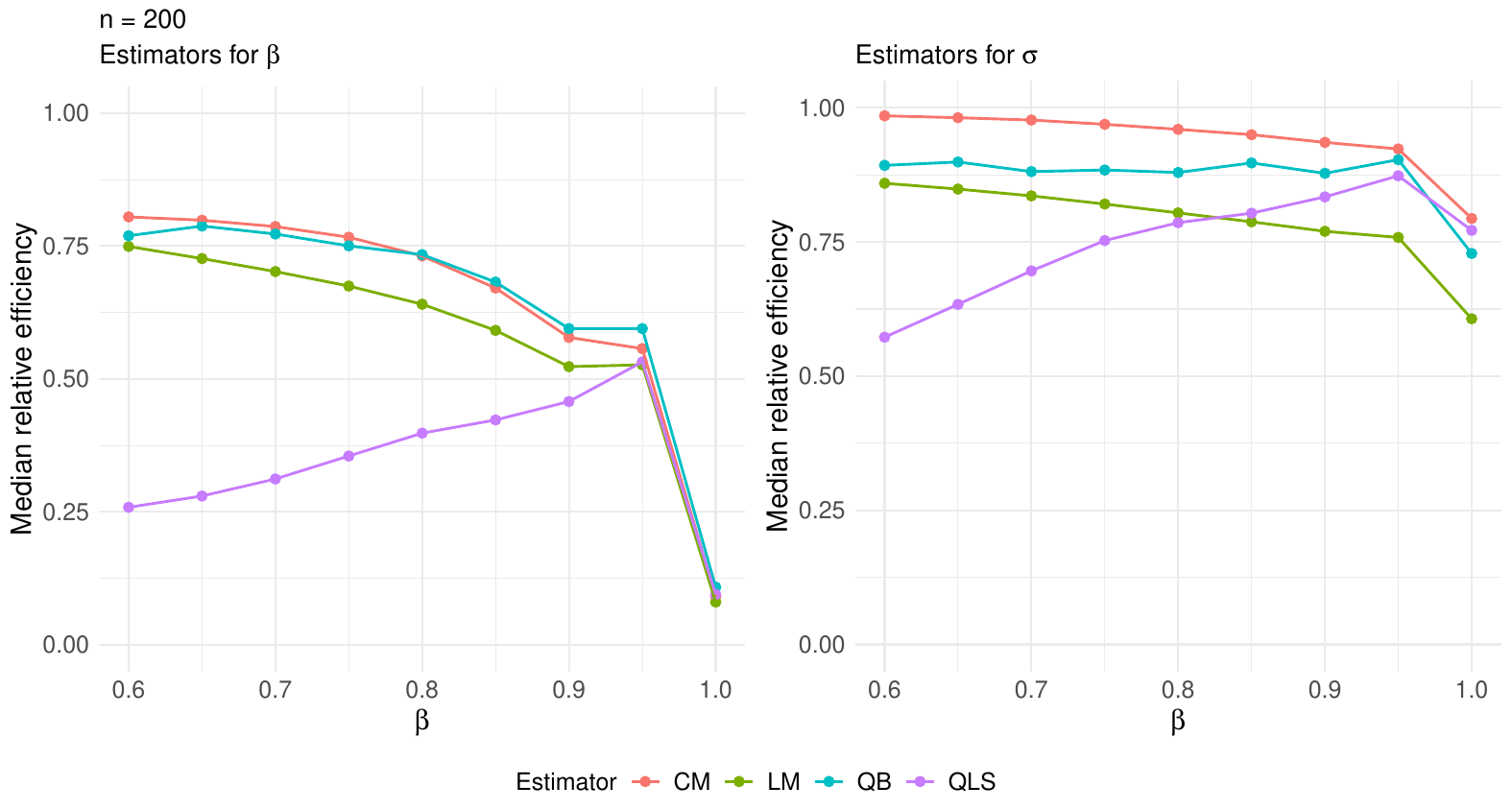}
    \caption{Median empirical relative efficiency of the CM, LM, QB and QLS estimators for the tail parameter $\beta$ (left) and scale parameter $\sigma$ (right), evaluated across different values of the underlying tail parameter $\beta$, with $n = 200$.}
    \label{fig_efficiency2}
\end{figure}

\begin{figure}[ht]
    \centering
    \includegraphics[width=0.92\linewidth]{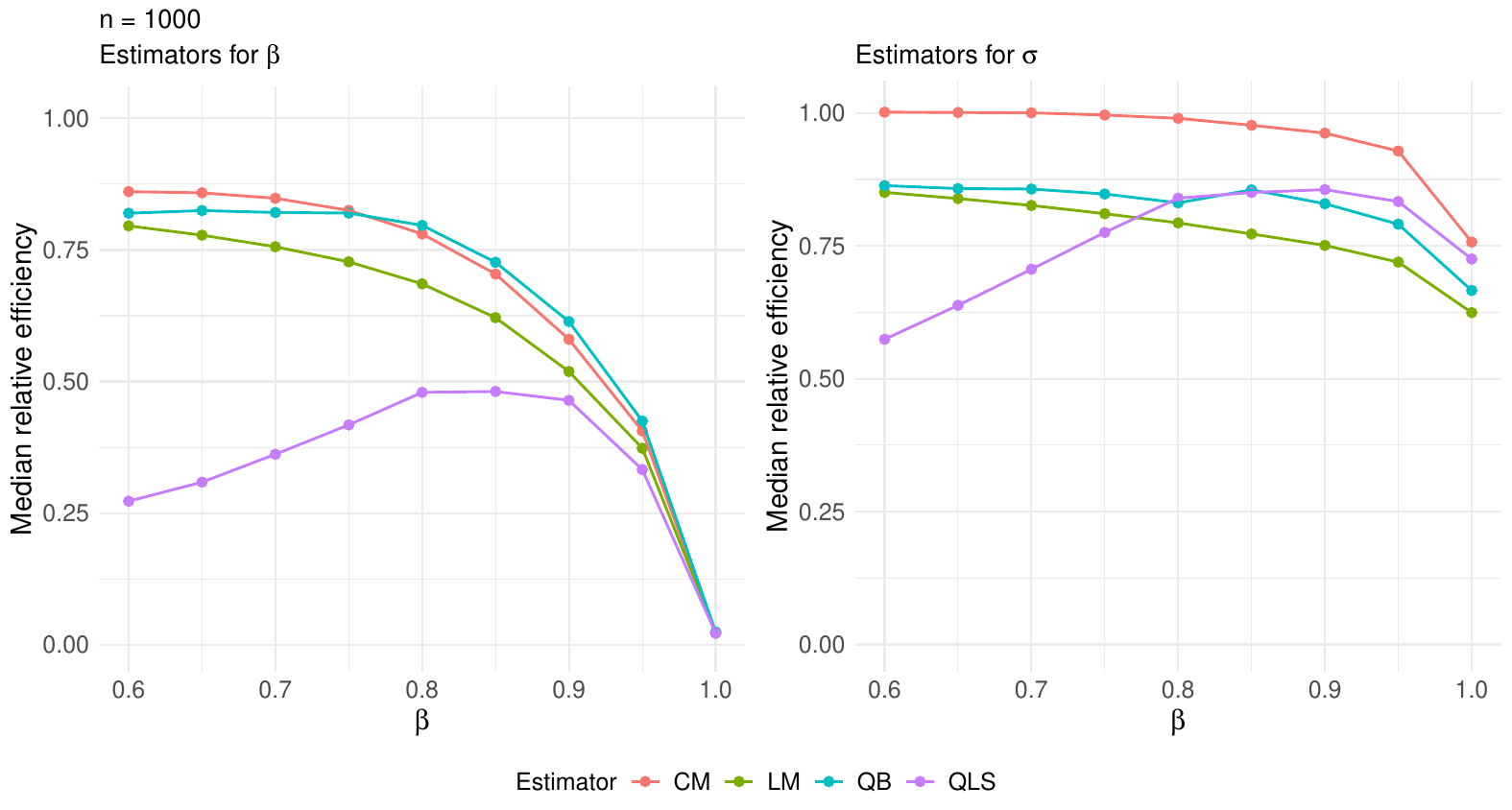}
    \caption{Median empirical relative efficiency of the CM, LM, QB and QLS estimators for the tail parameter $\beta$ (left) and scale parameter $\sigma$ (right), evaluated across different values of the underlying tail parameter $\beta$, with $n = 1000$.}
    \label{fig_efficiency3}
\end{figure}

\begin{figure}[!htbp]
    \centering
    \includegraphics[width=0.92\linewidth]{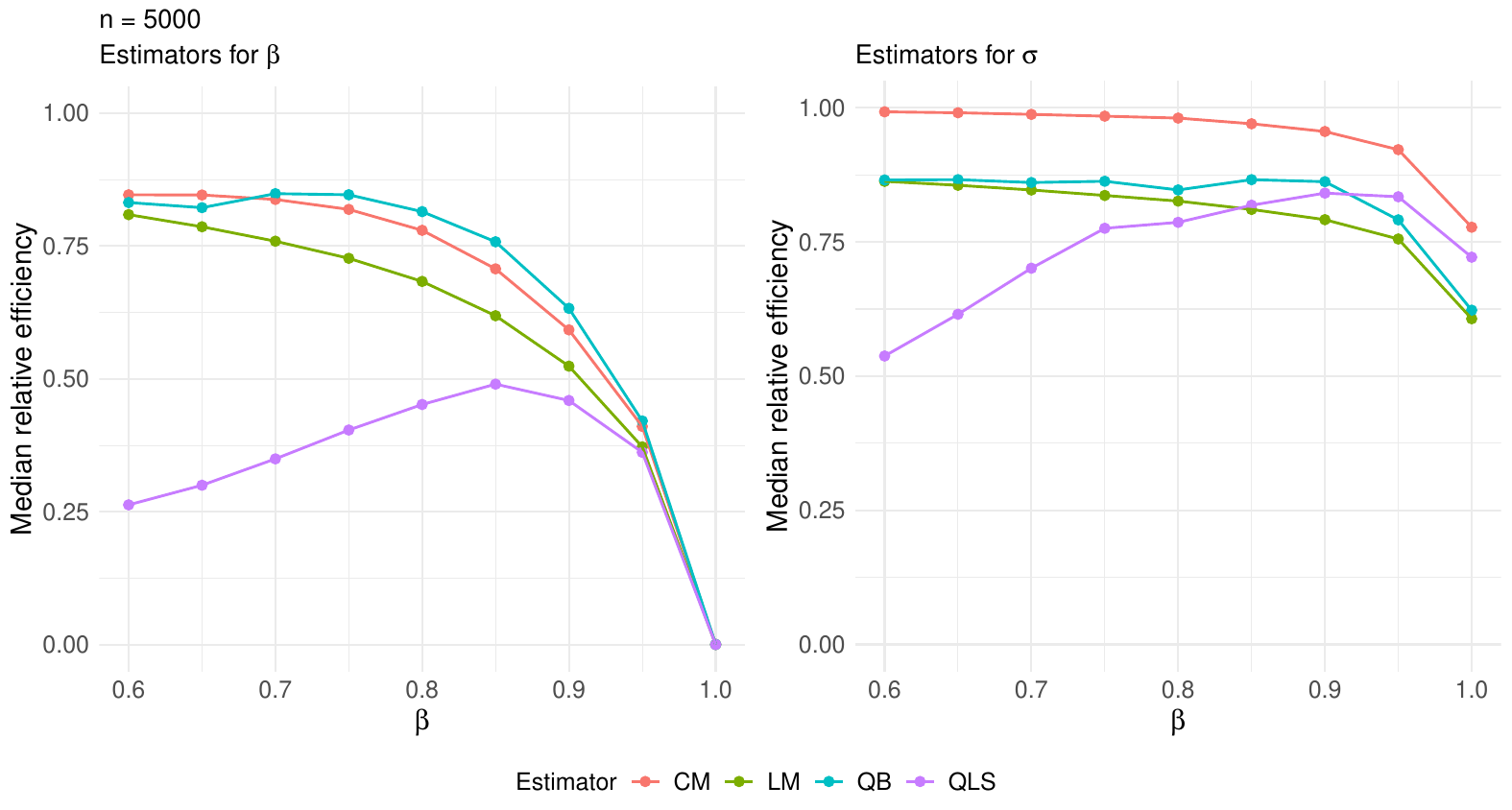}
    \caption{Median empirical relative efficiency of the CM, LM, QB and QLS estimators for the tail parameter $\beta$ (left) and scale parameter $\sigma$ (right), evaluated across different values of the underlying tail parameter $\beta$, with $n = 5000$.}
    \label{fig_efficiency4}
\end{figure}

\begin{figure}[!htbp]
    \centering
    \includegraphics[width=0.92\linewidth]{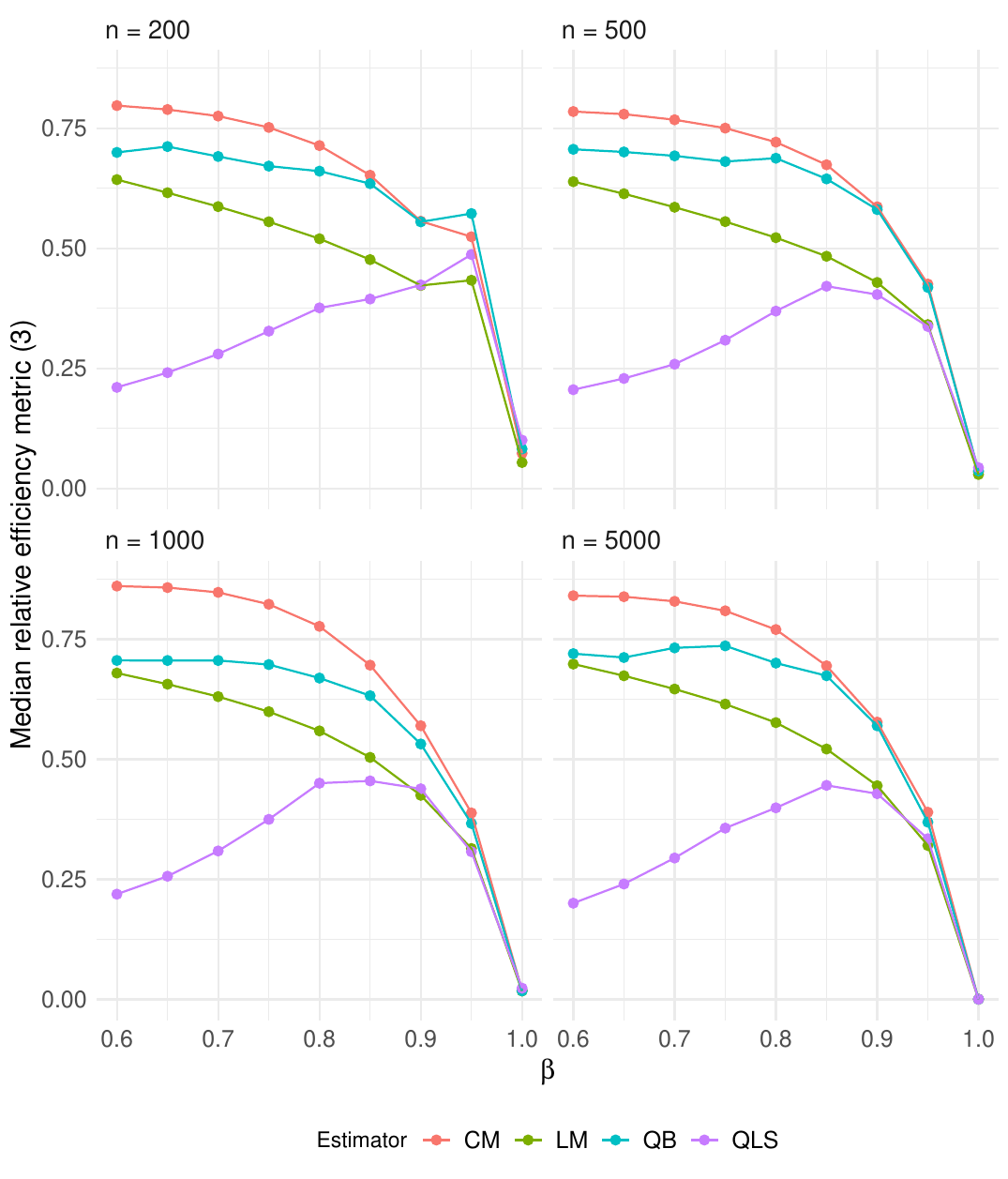}
    \caption{Median of the empirical relative efficiency metric (\ref{extra}) of the CM, LM, QB and QLS estimators across different values of the underlying tail parameter $\beta$, by sample size.}
    \label{fig_res_metric3}
\end{figure}

\end{document}